\documentclass[aps,prd,superscriptaddress,nofootinbib,eqsecnum, preprintnumbers]{revtex4}
\pdfoutput=1
\usepackage{amsmath, amsfonts, amssymb, latexsym, physics}
\usepackage{siunitx}
\usepackage{graphicx, subcaption}
\usepackage{placeins}
\usepackage{hyperref}
\usepackage{cleveref}

\newcommand{\dx}{D_{23}}
\newcommand{\dy}{D_{31}}
\newcommand{\dz}{D_{12}}

\begin{document}
\title{Stochastic gravitational wave background reconstruction\\ for a 
non-equilateral and unequal-noise LISA constellation}
\date{\today}
\author{Olaf Hartwig} \email{olaf.hartwig@obspm.fr}
\affiliation{SYRTE, Observatoire  de  Paris,  Universit\'e  PSL, CNRS,  Sorbonne  Universit\'e,  LNE,  61 avenue de l’Observatoire 75014  Paris,  France}
\author{Marc Lilley} \email{marc.lilley@obspm.fr }
\affiliation{SYRTE, Observatoire  de  Paris,  Universit\'e  PSL, CNRS,  Sorbonne  Universit\'e,  LNE,  61 avenue de l’Observatoire 75014  Paris,  France}
\author{Martina Muratore} \email{martina.muratore@aei.mpg.de}
\affiliation{Max-Planck-Institut für Gravitationsphysik (Albert-Einstein-Institut), \\ Am M\"uhlenberg 1, 14476 Potsdam, Germany}
\affiliation{Università degli studi di Trento, via Calepina, 14 - I-38122 Trento, Italy}
\author{Mauro Pieroni} \email{mauro.pieroni@cern.ch}
\affiliation{Department of Theoretical Phyics, CERN, 1211 Geneva 23, Switzerland}

\preprint{CERN-TH-2023-050}
\begin{abstract}%
We explore the impact of choosing different sets of Time-Delay Interferometry (TDI) variables for detecting and reconstructing Stochastic Gravitational Wave Background (SGWB) signals and estimating the instrumental noise in LISA. Most works in the literature build their data analysis pipelines relying on a particular set of TDI channels, the so-called AET variables, which are orthogonal under idealized conditions. By relaxing the assumption of a perfectly equilateral LISA configuration, we investigate to which degree these channels remain orthogonal and compare them to other TDI channels. We show that different sets of TDI variables are more robust under perturbations of the perfect equilateral configuration, better preserving their orthogonality and, thus, leading to a more accurate estimate of the instrumental noise. Moreover, we investigate the impact of considering the noise levels associated with each instrumental noise source to be independent of one another, generalizing the analysis from two to twelve noise parameters. We find that, in this scenario, the assumption of orthogonality is broken for all the TDI variables, leading to a misestimation of measurement error for some of the noise parameters. Remarkably, we find that for a flat power-law signal, the reconstruction of the signal parameters is nearly unaffected in these various configurations.
\end{abstract}
    
\maketitle
\tableofcontents
    
\section{Introduction}%
    
The Laser Interferometer Space Antella (LISA)~\cite{LISA:2017pwj} is a space mission led by the European Space Agency (ESA), also involving NASA, which is planned to be launched in the mid-2030s. LISA will consist of a constellation of three satellites separated by nearly 2.5 million kilometers, operating as a gravitational wave (GW) observatory in the milliHertz (mHz) range. LISA is expected to detect tens of thousands of resolvable sources during the expected 4-year mission duration, including Stellar Origin Binary Black Holes (SOBBHs), Compact Galactic Binaries (CGBs) comprised mostly of Double White Dwarfs (DWDs), Super Massive Black Holes (SMBHs) and Extreme Mass Ratio Inspirals (EMRIs). For a review of detection prospects for all these sources see, e.g.,~\cite{LISA:2022yao} and references therein. In addition to the resolvable sources enumerated above, a much larger number of weak and unresolvable sources will superimpose incoherently, leading to the generation of a Stochastic GW Background (SGWB). There are (at least) two guaranteed components contributing to the astrophysical SGWB in the LISA band\footnote{Depending on the detection rate, the SGWB due to EMRIs signal might, or might not, be detectable with LISA~\cite{LISA:2022yao}.}: at frequencies lower than few mHz, the dominant contribution will come from CGBs~\cite{1987ApJ...323..129E, Bender:1997hs}, while at higher frequencies another contribution is expected to originate from SOBBH mergers~\cite{LIGOScientific:2019vic}. Beyond astrophysical components, LISA could also be sensitive to cosmological SGWBs generated by violent processes taking place in the very early Universe. Detecting these signals would open a new window on energy scales beyond the reach of all the other probes used in particle physics/cosmology. Among the possible sources of cosmological SGWBs for LISA, let us mention inflation~\cite{Bartolo:2016ami}, cosmological Phase Transitions (PTs)~\cite{Caprini:2019egz}, Cosmic String (CS) networks~\cite{Auclair:2019wcv}, and second-order scalar induced tensor perturbations, typically associated with Primordial Black Holes production~\cite{LISACosmologyWorkingGroup_PBHs}. For reviews see, e.g.,~\cite{Caprini:2018mtu, LISACosmologyWorkingGroup:2022jok}.  One of the main challenges for SGWB detection is that these signals appear in the detector data stream as an additional noise source that has to be distinguished from the instrumental one. For this reason, SGWB detection and characterization requires dedicated methods~\cite{Allen:1997ad, Romano:2016dpx} that are quite different from the ones commonly employed for resolvable sources.\\

For what concerns instrumental noise sources, the main contribution in LISA will come from laser frequency noise~\cite{LISA_performance}. This critical noise source needs to be suppressed by several orders of magnitude to allow any GW detections with LISA. This noise suppression will be achieved using an on-ground data processing technique called Time-Delay Interferometry (TDI)~\cite{Tinto:2020fcc}. The results of the TDI algorithm are synthesized data streams representing several laser-noise-free virtual interferometers. As shown in~\cite{Armstrong_1999, Prince:2002hp, Shaddock:2003bc, Shaddock:2003dj, Tinto:2003vj, Vallisneri:2005ji, Muratore:2020mdf, Muratore:2021uqj}, it is possible to form several TDI channels which can have different sensitivities to GW signals and instrumental noise. The most commonly used TDI channels for LISA data analysis are the three Michelson like-variables, typically dubbed X, Y, and Z, often re-combined into the (quasi-)orthogonal channels, typically dubbed A, E, and T~\cite{Prince:2002hp}. Note that there are several motivations to advocate the search for orthogonal channels. For example, working in a diagonal TDI basis is convenient for data analysis algorithms. Indeed, since a diagonal matrix is trivial to invert, this avoids problems related to numerical inversion, thus improving the numerical stability and possibly significantly speeding up likelihood evaluations. \\

The AET basis is exactly orthogonal if LISA is a perfect equilateral triangle and if all secondary noise sources, i.e., noises that dominate the LISA data stream after laser noise suppression through TDI, such as Test Mass (TM) acceleration and Optical Metrology System (OMS) noise, are perfectly equal in all three spacecraft. In fact, for realistic motion of the satellites, the arm-lengths will be unequal (and time-varying) at the percent level~\cite{Martens:2021phh}, and LISA will not be a perfectly equilateral triangle. Moreover, the noises appearing in the different spacecraft will not be equal, breaking another assumption underlying the idealized derivation of the A, E, and T variables. Therefore the standard expressions for these channels will not be perfectly orthogonal when computed using actual LISA data\footnote{While, in principle, it is still possible to find other sets of TDI variables that form an orthogonal basis, this computation would have to be performed on the fly on the real data, vastly increasing the computational costs of the data analysis pipelines.}.\\

In this work, we first quantify the impact that a LISA configuration with unequal fixed armlengths and different noise levels among the satellites would have on the orthogonality of different sets of TDI variables by an explicit calculation of the noise and signal power spectral density (PSD) and cross-spectral density (CSD). For simplicity, we focus on the so-called first-generation TDI variables~\cite{Armstrong_1999}, which only fully suppress laser noise for a static LISA constellation, i.e., with arm-lengths that do not evolve in time.  We expect the main conclusions of this work to remain valid under our working assumptions for the second-generation variables~\cite{Hartwig:2021mzw}, which achieve laser noise suppression even for time-varying arm-lengths. Besides the most commonly used GW-sensitive Michelson variables XYZ and the corresponding quasi-orthogonal AET channels, we consider the Sagnac variables $\alpha \beta \gamma$ and their corresponding set of quasi-orthogonal channels, which we denote $\mathcal{A} \mathcal{E} \mathcal{T}$.  In addition, we consider the fully-symmetric Sagnac variable $\zeta$, which shares with T and $\mathcal{T}$ the property that it is quasi-insensitive to GWs and can be used instead of these channels to form a quasi-orthogonal set with A, E or $\mathcal{A}$, $\mathcal{E}$.\\

We then focus specifically on AET and AE$\zeta$ and use simulated data together with Markov Chain Monte Carlo (MCMC) Parameter Estimation (PE) as well as the Fisher Information Matrix (FIM) formalism to further study the sensitivity of both sets of TDI variables to a power law SGWB and to instrumental noise as defined in~\cite{LISA:2017pwj}, when cross-correlations are included or neglected. While both sets of variables are expected to be orthogonal sets and perfectly equivalent in the idealized situation of equal LISA arms and equal noise amplitudes, differences arise when the assumptions underlying the construction of the orthogonal channels are broken.\\
    
The paper is organized as follows. In~\cref{sec:measurement}, we describe the data model employed and derive the signal response of LISA to a SGWB for unequal LISA arms. We also show how the two dominant noise sources left after TDI (TM and OMS) appear in the LISA measurements. We then introduce the TDI formalism and derive a general formula for the noise and signal PSD and CSD for all the TDI channels considered in this work. In~\cref{sec:spectral-analysis}, we discuss the noise and signal PSDs and CSDs when the noise levels are assumed to be the same for all TM and OMS components, or when each TM and OMS component is assumed to be different. In either case, we compare the signal and noise correlations for equal and unequal LISA arms. In~\cref{sec:parameter-reconstruction}, as mentioned above, for a specific set of noise parameters and for a power-law SGWB signal, we produce simulated LISA data and perform PE using MCMC to compare the performance of two sets of TDI variables, namely AET and $AE\zeta$, when the cross-correlations in the TDI matrix are neglected.  Using FIM, we also study the impact of including or neglecting those cross-correlations on the PE results.  We do so for both equal and unequal noise levels and report on the results obtained for the SGWB in the main text, while those for the noise parameters can be found in an appendix.  We then conclude in~\cref{sec:conclusion}.\\

Our paper includes four appendices. \Cref{ssec:tdi-relations} discusses the relationships among the different sets of TDI variables considered. In \cref{sec:analytical-models}  we provide useful analytic approximations for the signal and noise spectra. In~\cref{sec:appendix_data_analysis}, we give an overview  of the data analysis method employed in~\cref{sec:parameter-reconstruction}. Finally, \cref{sec:noise_analysis} contains a detailed analysis of the noise reconstruction which serves to complement the SGWB signal parameter reconstruction provided in~\cref{sec:parameter-reconstruction}.

\section{Measurement characterization}%
\label{sec:measurement}               %
In this section, we describe the signal and noise components of the LISA data stream. After writing down a general model for the data as a superposition of signal and noise in~\cref{sec:data_model}, we derive the instrument response for an isotropic SGWB signal and the propagation of the different noise components in a single LISA link in~\cref{sec:signal_response} and~\cref{sec:noise_single_link}, respectively. In~\cref{sec:TDI_variables} we introduce the TDI variables that we will use in this paper. After providing the definitions of the base variables in~\cref{ssec:base-variables}, we introduce, in~\cref{ssec:tdi-csd-matrices}, a general method to compute the TDI PSDs and CSDs using the single link spectra computed in \cref{sec:signal_response} and~\cref{sec:noise_single_link}. 

\subsection{Data model \label{sec:data_model}}%
Let us start by assuming that all transient signals and glitches in the noise have been subtracted from the data stream, which is necessary to assume the noise to be stationary and Gaussian. For the moment, let us also restrict to the case of a single detector. Under these assumptions, the time domain data $d(t)$, can be expressed as a combination of the GW signal $s(t)$ plus the instrumental noise $ n(t)$ as:
\begin{equation}
d(t) = s(t) + n(t) \;  .
\end{equation}
While in reality, data will be sampled at a finite rate, in the following, we assume them to be continuous functions in the interval $[-T/2, T/2]$, with $T$ the observation time\footnote{This choice would only impact the high-frequency part of the data, where the frequency gets close to the Nyquist frequency. Since for LISA, the sampling rate is expected to be $\gtrsim$ 2Hz, and since in the analysis we only consider data up to $.5Hz$, this assumption should not impact the main results of this work.}. Assuming that signal and noise are uncorrelated, these quantities can be discussed separately. We start by discussing the signal properties, whose Fourier transform reads:
\begin{equation}
\tilde{s}(f) = \int_{-T/2}^{T/2} \textrm{e}^{2 \pi i f t } \; s(t) \; \textrm{d} t \; .
\end{equation}
Since $s(t)$ is real, $\tilde{s}$ obeys $\tilde{s}(f) = \tilde{s}^*(-f) $. Assuming stationarity, the expectation value of the signal's Fourier modes reads:
\begin{equation} \label{eq:psd-defintion}
\langle \tilde{s}(f)  \tilde{s}^*(f')  \rangle = \frac{1}{2} \delta(f- f') S^{\mathrm{GW}}(f) \; ,
\end{equation}
where $S^{\mathrm{GW}}(f)$ is a real and positive function with $S^{\mathrm{GW}}(f) = S^{\mathrm{GW}}(-f)$, which for a homogeneous and isotropic power spectrum\footnote{In general, the spectrum $P_{h}^{\lambda \lambda^\prime}$, with $\lambda$, $\lambda^\prime$ running over the two GW polarizations, defines a $2 \times 2$ matrix. The four entries of this matrix are typically expressed in terms of the Stokes parameters $I, V, Q, U$. In the L/R basis (see~\cref{eq:PC_LR_def}), $I$ and $V$, with $I$ the intensity and $V$ the circular polarization, only contribute to the diagonal while $Q$ and $U$ appear in the off-diagonal terms only. Homogeneity and isotropy correspond to vanishing $Q$ and $U$ Stokes parameters, implying that $P_{h}^{AB}$ is diagonal with $P_{h}^{RR} = I + V$ and  $P_{h}^{LL} = I - V$. Finally, for a non-chiral background, $V = 0$, which implies $ P_{h}^{RR} = P_{h}^{LL} = P_{h}$.} $P^{\lambda}_h$, with $\lambda$ denoting the two GW polarizations, see~\cref{eq:PC_LR_def} for the definition of the polarization tensors, can be expressed as:
\begin{equation} \label{eq:response-def1}
S^{\mathrm{GW}}(f)  = \sum_\lambda \mathcal{R}_\lambda (f) P^{\lambda}_h (f) \; ,
\end{equation}
where $\mathcal{R}_\lambda$, is the sky-averaged LISA response function. Note that since the LISA spacecraft lie in a plane, LISA cannot distinguish between chiralities without making use of the motion of the constellation~\cite{Seto:2007tn, Seto:2008sr, Smith:2016jqs, Domcke:2019zls}. Here, we assume the signal to be non-chiral i.e., $P^L_h  = P^R_h $, so that \cref{eq:response-def1} reduces to $S^{\mathrm{GW}}(f) = 2 \mathcal{R} (f) P_h (f)$. Absorbing the factor 2 into $\mathcal{R}$ and assuming the signal to be parity even (i.e., $P^{\lambda}(\vec{k}) = P(k)$) leads to
\begin{equation} \label{eq:response-def2}
S^{\mathrm{GW}}(f) = \mathcal{R} (f) \, P_h (f) \; .
\end{equation}
We conclude our discussion of the signal by recalling the expression of the signal power spectrum in units of the energy density parameter:
\begin{equation}
\Omega_{GW} h^2 \equiv  \frac{4 \pi^2  }{3 (H_0/h)^2} f^3 P_h(f) \; ,
\end{equation}
where $H_0 \approx 3.24 \times 10^{-18} h_0$ Hz is the Hubble constant today and $h_0 = 0.6766 \pm 0.0042$ is its dimensionless value~\cite{Planck:2018vyg}.\\
 
The expectation value of the noise's Fourier modes reads:
\begin{equation}
\langle \tilde{n}(f)  \tilde{n}^*(f')  \rangle = \frac{1}{2} \delta(f- f') S^\mathrm{N}(f) \; ,
\end{equation}
where we have introduced the noise power spectrum $S^\mathrm{N}(f)$, which satisfies the same properties as $S^{\mathrm{GW}}(f)$. Note that in the following sections, we consider the case of several data streams. In this generalized and more realistic scenario, the response function and noise spectra will be replaced by positive-definite Hermitian matrices. 
    
\subsubsection{Single link signal response \label{sec:signal_response}}%
To compute the GW response, we start by expressing a GW signal $h_{ab}(\vec{x}, t)$ as a superposition of plane waves~\cite{Maggiore:1999vm,Smith:2016jqs,Romano:2016dpx} (note that $c=1$):
\begin{equation}
h_{ab}(\vec{x}, t) = \int_{-\infty}^{\infty} \dd f \int \dd\Omega_{\hat{k}} \;  \textrm{e}^{2\pi i f(t- \hat{k}\cdot \vec{x})} \sum_A \tilde{h}_A(f,\hat{k})e^A_{ab}(\hat{k}) \; ,
\label{eq:h_frequency_expansion}
\end{equation}
with $f$ the GW frequency, $\hat{k}$ the outward vector in the direction of the incoming GW, $d\Omega_{\hat{k}}$ the infinitesimal solid angle and $e^A_{ab}(\hat{k})$ the polarization tensors. Following the convention of~\cite{Domcke:2019zls,Bartolo:2018qqn,Flauger:2020qyi}, given the normalized wave-vector $\hat{k}$, we can introduce the two vectors:
\begin{equation}
\hat{u}(\hat{k}) \equiv \frac{\hat{k} \times \hat{\rm e}_z}{|\hat{k} \times \hat{\rm e}_z|} \; , \qquad \qquad  \hat{v}(\hat{k}) \equiv \hat{k} \times \hat{u} \; , 
\end{equation}
where $\times$ denotes the external product and $\hat{\rm e}_z$ is the z-component vector of an arbitrarily oriented reference system. The $+/\times$ and $L/R$ polarization tensors are defined in terms of $ \hat{u}(\hat{k})$ and $\hat{v}(\hat{k})$ as:
\begin{equation}
\label{eq:PC_LR_def}
e^{+}_{ab}(\hat{k}) \equiv \hat{u}_a \hat{u}_b - \hat{v}_a \hat{v}_b \; , \qquad e^{\times}_{ab}(\hat{k}) \equiv \hat{u}_a \hat{v}_b + \hat{v}_a \hat{u}_b  \; , \qquad e^{L/R}_{ab}(\hat{k}) \equiv e^{+}_{ab}(\hat{k}) \mp i e^{\times}_{ab}(\hat{k})  \,,
\end{equation}
expressed in the L/R or $+/\times$ bases. We proceed by assuming that the LISA constellation is static and in a flat background spacetime. The time delay induced by GWs on a photon leaving at time $t - L_{ij}$ ($L_{ij}$ being the distance $|\vec{x}_i - \vec{x}_j|$) from $\vec{x}_j$ and reaching $\vec{x}_{i}$ at time $t$, can be expressed,  at lowest order in $h_{ab}$, as:
\begin{equation}
\Delta t_{ij}(t) \simeq \int_{0}^{L_{ij}} \frac{\hat l^a_{ij} \; \hat l^b_{ij}}{2} h_{ab}(t(s),\vec{x}(s))\; \dd s \; ,
\label{eq:etaij_start}
\end{equation}
where $\hat l_{ij} = (\vec x_j - \vec x_i)/|\vec x_j - \vec x_i|$ is a unit vector pointing from $i$ to $j$, and $t(s) \equiv t - L_{ij} + s$, $\vec x(s) = \vec x_j - s\, \hat l_{ij}$ are respectively the time and position along the photon path expressed in terms of the affine parameter $s$. By inserting~\cref{eq:h_frequency_expansion} into~\cref{eq:etaij_start}, and by considering the fractional frequency shift rather than the time delay induced by the GWs, we obtain:
\begin{equation}
\eta_{ij}^{\mathrm{GW}}(t) = \frac{\dd}{\dd t}\Delta t_{ij}(t) =  i \int_{-\infty}^{\infty} \dd f\, \frac{f}{f_{ij}}\textrm{e}^{2\pi i f(t - L_{ij})}
\int \dd\Omega_{\hat{k}} \left[ \textrm{e}^{-2\pi i f \hat{k}\cdot \vec x_i } \sum_A \xi^A_{ij}(f, \hat k)\tilde{h}_A(f,\hat{k})]\right] \; ,
\label{eq:etaij_final}
\end{equation}
where we have introduced the characteristic frequencies $f_{ij} \equiv (2 \pi L_{ij})^{-1}$ and the functions $\xi^A_{ij}\left(f, \hat k\right)$ defined as:
\begin{equation}
\xi^A_{ij}\left(f, \hat k\right)= e^{-2\pi i f \hat k\cdot \vec L_{ij}}\mathcal M_{ij}(f, \hat k) \; \mathcal{G}^A(\hat k,\hat l_{ij}) \; ,
\end{equation}
with:
\begin{equation}
\mathcal{M}_{ij}(f, \hat k) \equiv \mathrm{e}^{\pi i f L_{ij} ( 1 + \hat{k}\cdot \hat l_{ij})}  \; \mathrm{sinc}\left(\pi f L_{ij} ( 1 + \hat{k}\cdot \hat l_{ij})\right) \quad \mathrm{and} \quad\mathcal{G}^A(\hat k, \hat l_{ij}) \equiv \frac{\hat l^a_{ij}\hat l^b_{ij}}{2}e^A_{ab}(\hat k) \; .
\end{equation}
Using~\cref{eq:psd-defintion}, the CSD for $\tilde{\eta}_{ij}^{\mathrm{GW}}(f)$, i.e. the Fourier transform of~\cref{eq:etaij_final}, then reads\footnote{This equation assumes $T f \gg 1$, so that the finite-time delta functions, arising from the Fourier transform, can be replaced with real delta functions.}:
\begin{equation}
S^{\eta,\mathrm{GW}}_{ij,mn}(f) \equiv \sum_A \mathcal{R}^A_{ij,mn} \, P_h^{AA}(f)  = \frac{f^2}{f_{ij}f_{mn}} e^{-2\pi i f(L_{ij}-L_{mn})}\sum_A P_h^{AA}(f)  \; \Upsilon_{ij,mn}^{A}(f) \; ,
\label{eq:CSDsignal}
\end{equation}
with:
\begin{equation}
\Upsilon_{ij,mn}^{A}(f)=\int \frac{\dd\Omega_{\hat{k}}}{4 \pi}  \; \textrm{e}^{-2\pi i f \hat{k}\cdot (\vec x_i - \vec x_m)}  \;  \xi^A_{ij}(f, \hat k)  \, \xi^A_{mn}(f, \hat k)^* \; ,
\label{eq:Upsilon}
\end{equation}
where we have used the statistical properties of $\langle h_A(f,\hat k) \, h_B^*(f', \hat k')\rangle$ for a homogeneous, isotropic and non-chiral SGWB, i.e.:
\begin{equation} \label{eq:h-statistics}
\langle \tilde h_A(f,\hat k) \, \tilde h_B^*(f', \hat k')\rangle = \delta(f - f')\delta(\hat k -\hat k')\delta_{AB}\frac{P_{h}^{AB}(f)}{16\pi} \; \qquad \qquad  \langle \tilde h_A(f,\hat k) \, \tilde h_B(f',\hat k')\rangle = 0 \; ,
\end{equation}
with $P_{h}^{AB}(f)$ the one-sided PSD. The integral appearing in~\cref{eq:CSDsignal} can generally be computed numerically and in the low-frequency approximation, it can be computed analytically (see~\cref{sec:analytical-models}, in particular~\cref{sec:signal_resp_low_f}). 
    
\subsubsection{Single link noise spectra}%
\label{sec:noise_single_link}
In this section we discuss the two main unsuppressed secondary noises limiting the performance of LISA. We consider TM acceleration noise, and OMS noise~\cite{LISA:2017pwj}. These noise sources enter the single link measurement as\footnote{For a more detailed computation considering the ``split interferometry'' scheme see~\cite{Hartwig:2021dlc}. Note that the TM and OMS noise contributions have different correlation properties, which will ultimately cause the two components to have different transfer functions in the various TDI channels.}:
\begin{equation}
\eta_{ij}^\mathrm{N}(t) = n^\text{OMS}_{ij}(t) +  D_{ij} n_{ji}^\text{TM}(t) + n_{ij}^\text{TM}(t) \; , \label{eq:eta}
\end{equation}
 where $D_{ij}$ is the delay operator, which in the static LISA arm approximation used in this work acts on any time-dependent function $x(t)$ as $D_{ij}x(t) = x(t - L_{ij})$. We formally define the single links noise CSDs as:
\begin{equation}
\langle \tilde \eta_{ij}^\mathrm{N}(f) \, \tilde \eta_{lm}^{\mathrm{N}*}(f') \rangle = \frac{1}{2} \, S_{ij,lm}^{\eta,\mathrm{N}}(f) \, \delta(f - f') \; ,\label{eq:csd_def}
\end{equation}
and assume the individual noise terms to be stationary, zero mean, and uncorrelated, such that cross-terms between noises vanish, with the only non-zero terms given by:
\begin{subequations}
\begin{align}
\langle \tilde n^\text{OMS}_{ij}(f) \, \tilde n^\text{OMS*}_{ij}(f') \rangle &= \frac{1}{2}  \, S^\text{OMS}_{ij}(f) \,\delta(f - f') \; , \\
\langle \tilde n^\text{TM}_{ij}(f) \, \tilde n^\text{TM*}_{ij}(f') \rangle &= \frac{1}{2} \, S^\text{TM}_{ij}(f) \, \delta(f - f') \; .
\end{align} \label{eq:csd_noise}
\end{subequations}
Here, $S^\text{OMS}_{ij}(f)$ and $S^\text{TM}_{ij}(f)$ are the PSDs of the individual OMS and TM acceleration noise terms. Considering the definition of the noise CSD and the single link measurement, given~\cref{eq:eta} and~\cref{eq:csd_noise}, we can then directly compute the non-zero entries of the noise CSD of the single links as:
    \begin{subequations}\label{eq:csd_noises}
        \begin{align}
            S_{ij,ij}^{\eta,\mathrm{N}}(f) &= S^\text{OMS}_{ij}(f) + S^\text{TM}_{ij}(f) + S^\text{TM}_{ji}(f) \; ,  \\
            S_{ij,ji}^{\eta,\mathrm{N}}(f) &= e^{2\pi i f L_{ji}}S^\text{TM}_{ij}(f) +e^{-2\pi i f L_{ij}} S^\text{TM}_{ji}(f) \; .
        \end{align}
    \end{subequations}
In the following, we will further assume all noises of the same type to have the same spectral shape given by~\cite{LISA:2017pwj}:
\begin{subequations}
\begin{align}
\label{eq:Acc_noise_def}
S^\text{TM}_{ij}(f) &= A_{ij}^2 \times 10^{-30} \;  \times \qty(1 + \qty(\frac{\SI{0.4}{\milli\hertz}}{f})^2)\qty(1 + \qty(\frac{f}{\SI{8}{\milli\hertz}})^4)\times \qty(\frac{1}{2 \pi f c})^2 \; \times ( \textrm{m}^2 / \textrm{s}^3 ) \;,  \\ 
\label{eq:OMS_noise_def}
S^\text{OMS}_{ij}(f) &= P_{ij}^2 \times 10^{-24} \;  \times\qty(1 + \qty(\frac{\SI{2e-3}{\hertz}}{f})^4 ) \times \qty(\frac{2 \pi f}{c})^2 \; \times ( \textrm{m}^2 / \textrm{Hz} ) \;, 
\end{align}
\end{subequations}
such that each noise depends only on a single constant amplitude parameter $A_{ij}$ for TM and $P_{ij}$ for OMS. These parameters are dimensionless so that the overall noise PSDs are given in units of fractional frequency deviations. \\ In this work, we consider two scenarios: 
\begin{enumerate}
\item The noise amplitudes of TM on the one hand and OMS on the other are equal, i.e., $A_{ij} = A$ and $P_{ij} = P$, with central values respectively given by $A = 3 $ and $P = 15$~\cite{Babak:2021mhe}.
\item The noise parameters $A_{ij}$  and $P_{ij}$ have random values lying within a standard deviation of $20\%$ around the central values\footnote{Randomly drawing the noise amplitudes with a standard deviation of $20\%$ causes the largest and smallest noise terms to be approximately within a factor 2. This is roughly in line with what was observed in LISA PathFinder (LPF), where the observed noise levels were within a factor of a few of their anticipated values~\cite{Castelli:2020zro}. Note that while the LPF noise measured in flight did not agree with the predicted noise level nor with its {\it shape} (especially at frequencies below $10^{-3}$ Hz), here, we make the rather strong assumption that the noise {\it shapes} are perfectly known.}. The exact values of the $A_{ij}$ and $P_{ij}$ (with $\left\{ ij \right\} \in \mathcal{I} = \{12,23,31,21,32,13\}$) are given in~\cref{sec:un_noise}.
\end{enumerate}
    
\begin{figure}[t]
\centering
\includegraphics[width=.5\linewidth]{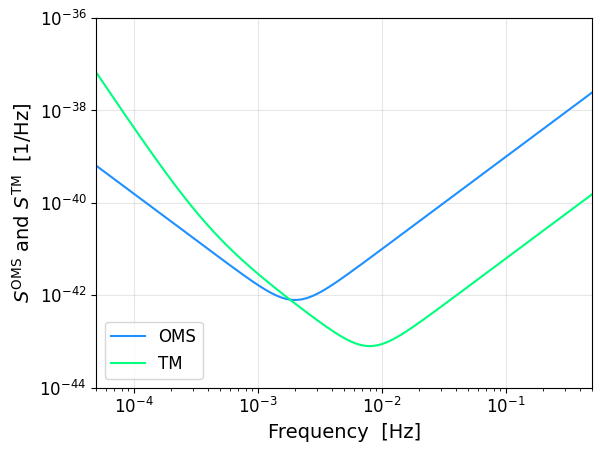}
\caption{TM and OMS PSDs as a function of frequency in units of $1/$Hz.}
\label{fig:noise_psds}
\end{figure}

For reference, a plot of the noise levels for the case $A_{ij}=A=3$ and $P_{ij}=P=15$, is shown in~\cref{fig:noise_psds}.
    
\subsection{TDI variables} %
\label{sec:TDI_variables}  %
The TDI variables that we will consider throughout this paper are based on the standard  X, Y, and Z Michelson combinations, as well as the $\alpha$, $\beta$, and $\gamma$~\cite{Armstrong_1999} Sagnac variables. We will also discuss the orthogonal channels A, E, and T and $\mathcal{A}$, $\mathcal{E}$ and $\mathcal{T}$ build out of these sets of base variables, respectively, as well as the fully symmetric Sagnac (null channel) $\zeta$~\cite{Hogan:2001jn,Muratore:2021uqj,Tinto:2020fcc}. For simplicity, we consider the first-generation versions of all these variables, which fully suppress laser noise only for a constellation with three unequal but {\it constant} arms\footnote{Laser noise suppression with more realistic orbits requires additional ``virtual loops'' in the photon paths that make up the TDI combinations. Following~\cite{Vallisneri:2005ji,Muratore:2020mdf,Muratore:2021uqj,Hartwig:2021mzw}, there exist several second-generation versions of each of the first-generation TDI variables. The ``standard'' choices that are found in the literature can be approximated as:
\begin{subequations}
\begin{align}
{\rm X}_2 &= (1 - D_{31}^2 D_{12}^2) {\rm X} \; , \\
\alpha_2 &= (1 - D_{12} D_{23} D_{31}) \alpha \; ,\\
\zeta_2 &= (D_{31} - D_{12}D_{23}) \zeta \; .
\end{align}
\end{subequations}
Note that $\zeta_2$ here refers to the variable recently found in~\cite{Muratore:2020mdf}, not the traditional variable given in~\cite{Tinto:2003vj}, which is less effective at suppressing laser noise than other second-generation TDI variables.
These additional pre-factors should not significantly impact the sensitivity of the TDI variables taken individually, as they apply equally to both noise and signal. However, we remark that the factors applied to ${\rm X}_2$, ${\rm Y}_2$ and ${\rm Z}_2$ are different, such that the full CSD matrix of the second-generation Michelson variables cannot be trivially derived from the first-generation versions. It is different for the Sagnac variables, where all three permutations are modified by the same overall factor.}, and compute their signal and noise CSDs in terms of the single-link CSDs given in the previous sections.
    
\subsubsection{Base variable definition\label{ssec:base-variables}}%
Let us start by defining the first-generation version of the XYZ Michelson variables. The X variable is defined as:%
\begin{subequations}\label{eq:tdi-definition}
\begin{equation}
{\rm X}  = (1 - D_{13}D_{31})(\eta_{12} + D_{12} \eta_{21}) + (D_{12}D_{21} - 1)(\eta_{13} + D_{13} \eta_{31}) \; ,
\end{equation}
while Y and Z are cyclic permutations of X. The Sagnac variable $\alpha$ is defined as:
\begin{equation}
\alpha = \eta_{12} + D_{12}\eta_{23} + D_{12}D_{23}\eta_{31} - (\eta_{13} + D_{13}\eta_{32} + D_{13}D_{32}\eta_{21}) \; , \label{eq:alpha1}
\end{equation}
with again $\beta$, $\gamma$ defined as cyclic permutations of $\alpha$. Finally, the fully symmetric $\zeta$ variable is defined as:
\begin{equation}
\zeta = D_{12}(\eta_{31} - \eta_{32}) + D_{23}(\eta_{12} - \eta_{13}) + D_{31}(\eta_{23} - \eta_{21}) \; .
\label{eq:zeta-1.5}
\end{equation}
\end{subequations}
Following~\cite{Prince:2002hp}, we introduce the so-called quasi-orthogonal TDI channels, which are usually given as:
\begin{equation}
\mathcal{A} = \frac{\gamma - \alpha}{\sqrt{2}} \;, \qquad  \mathcal{E} = \frac{\alpha - 2 \beta + \gamma}{\sqrt{6}} \;, \qquad  \mathcal{T} =\frac{\alpha + \beta + \gamma}{\sqrt{3}} \; ,
\end{equation}
for the Sagnac variables. An analogous procedure can be carried out for the Michelson variables, giving:
\begin{equation}
{\rm A} = \frac{{\rm Z} - {\rm X}}{\sqrt{2}}\;, \qquad  {\rm E} = \frac{{\rm X} - 2 {\rm Y} + {\rm Z}}{\sqrt{6}} \;, \qquad  {\rm T}=\frac{{\rm X} + {\rm Y} + {\rm Z}}{\sqrt{3}}  \; .
\end{equation}
These channels are designed to be orthogonal for both signal and noise, at least in the idealized case of equal arm and equal and symmetric noise levels in the three base variables. Having orthogonal channels drastically simplifies the computation of the inverse of the covariance matrix appearing in the likelihood, which makes these channels attractive for the practical application of LISA data analysis. We will discuss in~\cref{sec:spectral-analysis} the extent to which the orthogonality of these channels survives if one relaxes some of the assumptions used to derive them. Note finally that the different TDI variables discussed here are not independent, see~\cref{ssec:tdi-relations} for more details.

\subsubsection{Signal and noise projection on the TDI variables \label{ssec:tdi-csd-matrices}}%
In order to compute the noise PSDs and the response to a SGWB, we first need to evaluate the Fourier transform of any TDI variable $V$, for which we use the compact vector notation:
\begin{equation}
\tilde V(f) = \sum_{ij\in \mathcal I}  \, c^V_{ij}(f) \; \tilde{\eta}_{ij}(f) \; ,  \label{eq:Vf}
\end{equation}
where, as in~\cref{sec:noise_single_link}, $\mathcal{I} = \{12,23,31,21,32,13\}$ denotes the pairs of indices that define the six inter-satellite links and where the coefficients $c^V_{ij}$ map the single-link measurements onto the TDI variable $V$. Since we work in the assumption of constant delays, we can directly read off the coefficients $c^V_{ij}$ from~\cref{eq:tdi-definition} by replacing each delay in the time domain, $D_{ij}$, with the corresponding frequency domain expression $e^{-i 2 \pi f L_{ij}}$. Assuming $\tilde{U}$ and $\tilde{V}$ to be any two TDI variables, which, similarly to~\cref{eq:csd_def}, obey:
\begin{equation}
\langle \tilde{U}(f)  \, \tilde{V}^*(f')\rangle = \frac{1}{2} \, S^{UV}(f)  \, \delta(f - f') \;, 
\end{equation}
we can substitute~\cref{eq:Vf} into this expression to get:
\begin{subequations}\label{eq:TDI-cross-spectra}
\begin{align} 
\langle \tilde{U}(f) \tilde{V}^*(f')\rangle
&= \sum_{ij,mn\in \mathcal{I}} c^U_{ij}(f) \, c^{V*}_{mn}(f')\langle \tilde\eta_{ij}(f) \,  \tilde\eta_{mn}^*(f')\rangle \; , \\
&= \frac{1}{2}\underbrace{\sum_{ij,mn\in \mathcal{I}} \underbrace{c^U_{ij}(f) \,  c^{V*}_{mn}(f)}_{C^{UV}_{ij,mn}(f)} \,  S^\eta_{ij,lm}(f)}_{S^{UV}(f)} \,  \delta(f - f') \; .
\end{align}
\end{subequations}
For any particular $(\tilde{U}\,,\tilde{V})$ combination, the coefficients $c^U_{ij}(f)$ and $c^V_{ij}(f)$ can be combined to form a $6\times6$ matrix\footnote{A compact notation one could use in order to consider several, say $n$, TDI variables would require for the coefficients to be arranged into $n \times 6$ matrices, and the $C^{UV}$ to be replaced by a rank 4 tensor ($n \times n \times 6 \times 6 $), which maps the $ 6 \times 6$ single-link correlations onto the $ n \times n$ TDI variable correlations.} $C^{UV}$. We remark that this procedure is similar to what has very recently (and independently from our work) been proposed in~\cite{Baghi:2023qnq}.
Note that the coefficient matrices $C^{UV}$ depend only on the choice of TDI variables, while the actual noise or signal correlations are encoded in the previously computed single-link correlation matrix $S^\eta$. In the following sections, we will use:
\begin{equation}
\label{eq:signal_and_noise_projections}
S^{UV,\mathrm{N}} = C^{UV} S^{\eta,\mathrm{N}},\quad S^{UV,\mathrm{GW}} = C^{UV} S^{\eta,\mathrm{GW}} \; ,
\end{equation}
to identify the noise and signal TDI covariance, respectively.  Explicit expressions for the CSDs are given in~\cref{sec:analytical-models}.
    
\section{Relaxing the equilateral assumption: spectral analysis}%
\label{sec:spectral-analysis}

In this section, we present the signal response, the noise spectra, and the GW sensitivities for all the TDI variables considered in this work. We consider both the case of an equilateral and that of a non-equilateral LISA configuration. We perform a detailed correlation analysis to test the robustness of the orthogonalization procedure for the different TDI bases. The section is divided into two parts: the first part concerns the case of equal noise levels while the second part treats the case of unequal noise levels. All plots presented in this section will show the equal-arm model with solid lines while the unequal-arm results are plotted using dashed lines. Analytic expressions for the noise CSDs in the equal and (in the low-frequency limit) unequal arms case are provided in~\cref{sec:analytical-models}. The full expressions for the noise spectra are provided as supplementary material. The signal response shown in the plots was evaluated numerically (analytic expressions for the signal exist only in the low-frequency approximation, which are also provided in~\cref{sec:analytical-models}).

\begin{figure}[htb!]
\centering
\includegraphics[width=\linewidth]{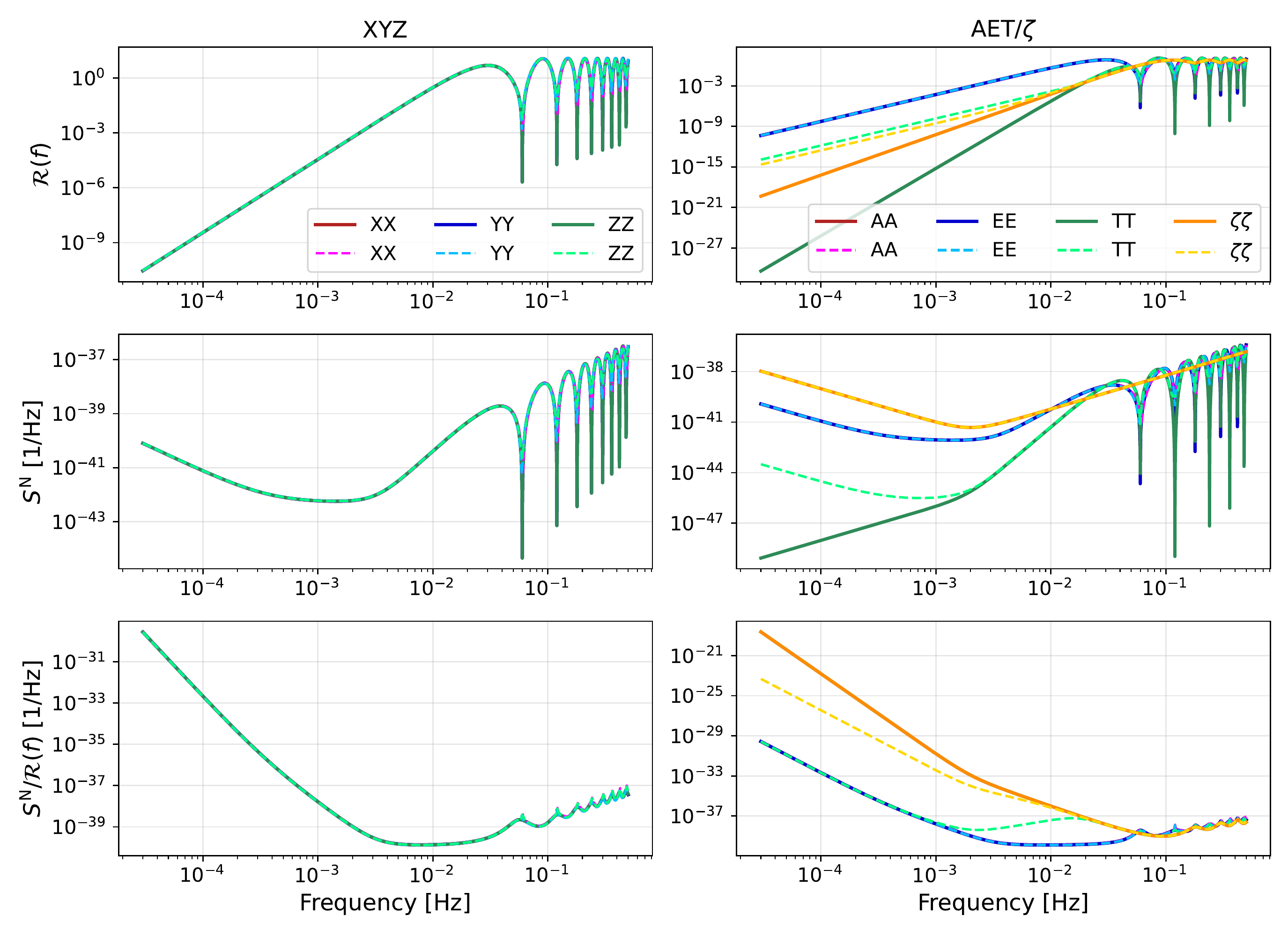}
\caption{Signal response (first row), noise spectra (second row), and strain sensitivities (last row) defined as in~\cref{eq:signal_and_noise_projections}, for the XYZ Michelson variables (first column) and for their orthogonal AET combinations as well as for the $\zeta$ variable (second column) considering equal (solid lines) and un-equal (dashed-line) arm-lengths.}
\label{fig:signal_psds_XYZ}
\end{figure} 

\subsection{Equal noise levels}%
\label{ssec:self-correlations-equal}

Let us focus on the case where all the noises of the same type are characterized by a single amplitude parameter, i.e., $A_{ij} \equiv A$, $P_{ij} \equiv P$ so that $S_{ij}^{\mathrm{OMS}} \equiv S^{\mathrm{OMS}}$ and $S_{ij}^{\mathrm{TM}} \equiv S^{\mathrm{TM}}$ for all links. \Cref{fig:signal_psds_XYZ} shows the three quantities that best describe the self-correlations of the XYZ Michelson variables, the AET orthogonal variables, and as well as the $\zeta$ channel. In the top row, we plot the quadratic signal response $\mathcal R$, as defined in~\cref{sec:data_model}, in units of squared fractional frequency deviation. We observe that while the equal and unequal arms models agree with one another for the signal-dominated channels A and E, they disagree at low frequencies for the null channels T and $\zeta$, especially for T. In the second row, we show the full noise spectra, $S^{UU,N}$, in units of squared fractional frequency deviation per \si{\hertz}. We notice that the equal and unequal arms models agree for all channels with the exception of T. Finally, the third row shows the strain sensitivity\footnote{Note that the strain sensitivity is a function of the noise model, the instrument response, and the TDI variable, but is independent of the GW signal.} computed as the ratio between the noise PSD $S^{UU,N}$ and the signal response $\mathcal{R}$. We observe that the strain sensitivities of X, Y, Z, A, and E are well approximated by the equal arm model, while there are sizeable differences for the two null channels. T, in particular, is no longer a null channel below \SI{1}{\milli\hertz}, and becomes as sensitive to the signal as A and E (as also reported in~\cite{Adams:2010vc} and more recently in~\cite{Muratore:2021uqj}). Notice that $\zeta$ also becomes more sensitive to GWs, but remains significantly less sensitive than A and E. \\

\begin{figure}[htb!]
\centering
\includegraphics[width=\linewidth]{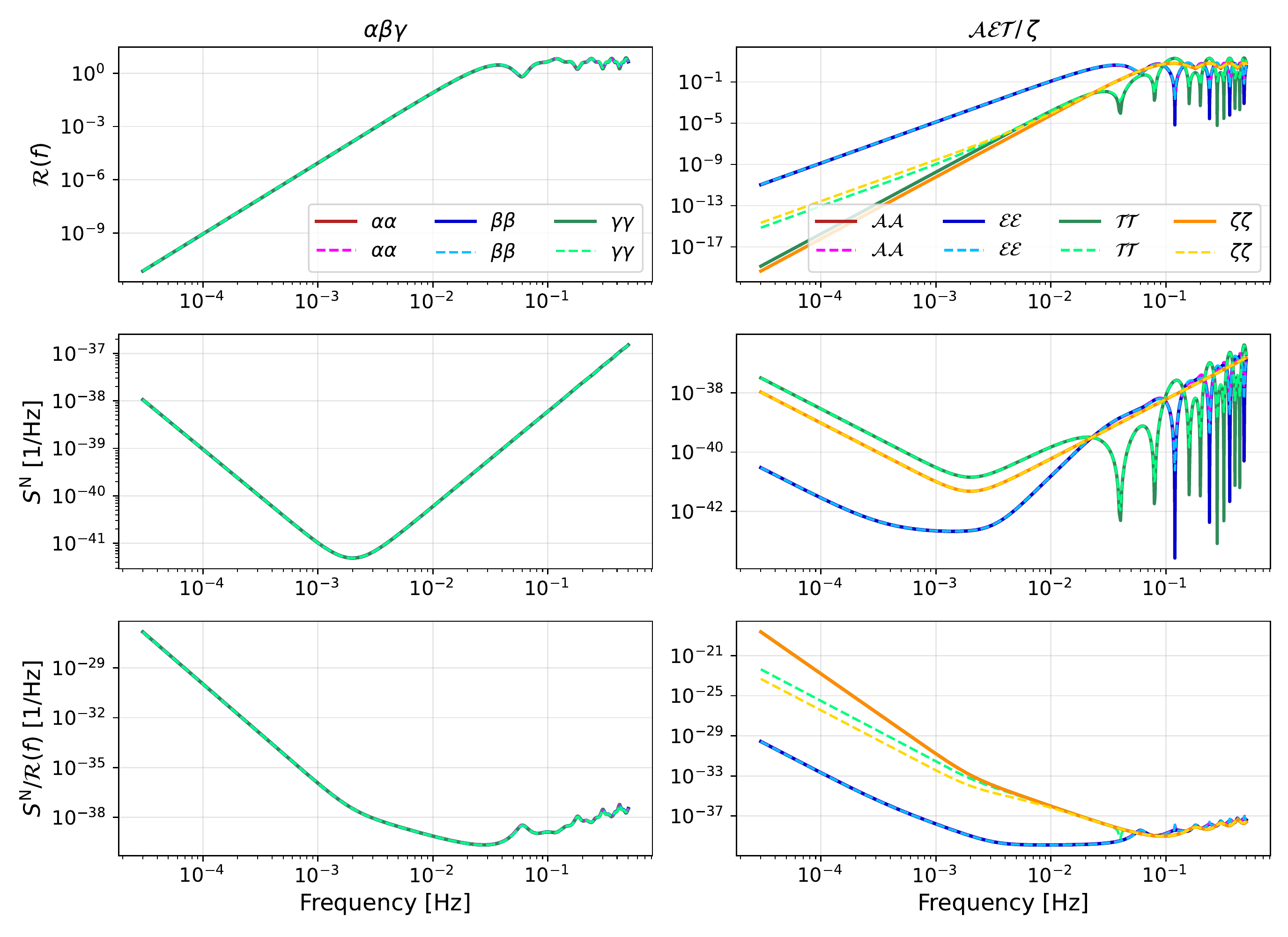}
\caption{Signal response (first row), noise spectra (second row) and strain sensitivities (last row) defined as in~\cref{eq:signal_and_noise_projections},  for the Sagnac $\alpha, \beta, \gamma$ variables (first column) and for their orthogonal combinations $\mathcal{A}$, $\mathcal{E}$, and $\mathcal{T}$ as well as for the $\zeta$ variable (second column).}
\label{fig:signal_psds_ABG}
\end{figure}

In~\cref{fig:signal_psds_ABG} we show the same quantities as in~\cref{fig:signal_psds_XYZ}, but for the Sagnac variables, the orthogonal variables $\mathcal{A}$, $\mathcal{E}$ and $\mathcal{T}$, and again the channel $\zeta$. Once again, the base variables $\alpha$, $\beta$, $\gamma$ as well as the sensitive channels $\mathcal{A}$, $\mathcal{E}$ are well approximated by the equal arms model. The null channel $\mathcal{T}$ behaves very similarly to $\zeta$, meaning that its noise spectrum is also well described using an equal arms model.  It remains a valid null channel at low frequencies and slightly outperforms $\zeta$ for unequal arms. Let us also note that the noise spectra of the base variables $\alpha$, $\beta$, $\gamma$, and $\zeta$ have no zeros within the LISA band. While this property is not fully retained by $\mathcal{A}$ and $\mathcal{E}$, they do have fewer zeros than A and E (compare~\cref{fig:signal_psds_XYZ}).\\
    
\begin{figure}[t]
\centering
\includegraphics[width=\linewidth]{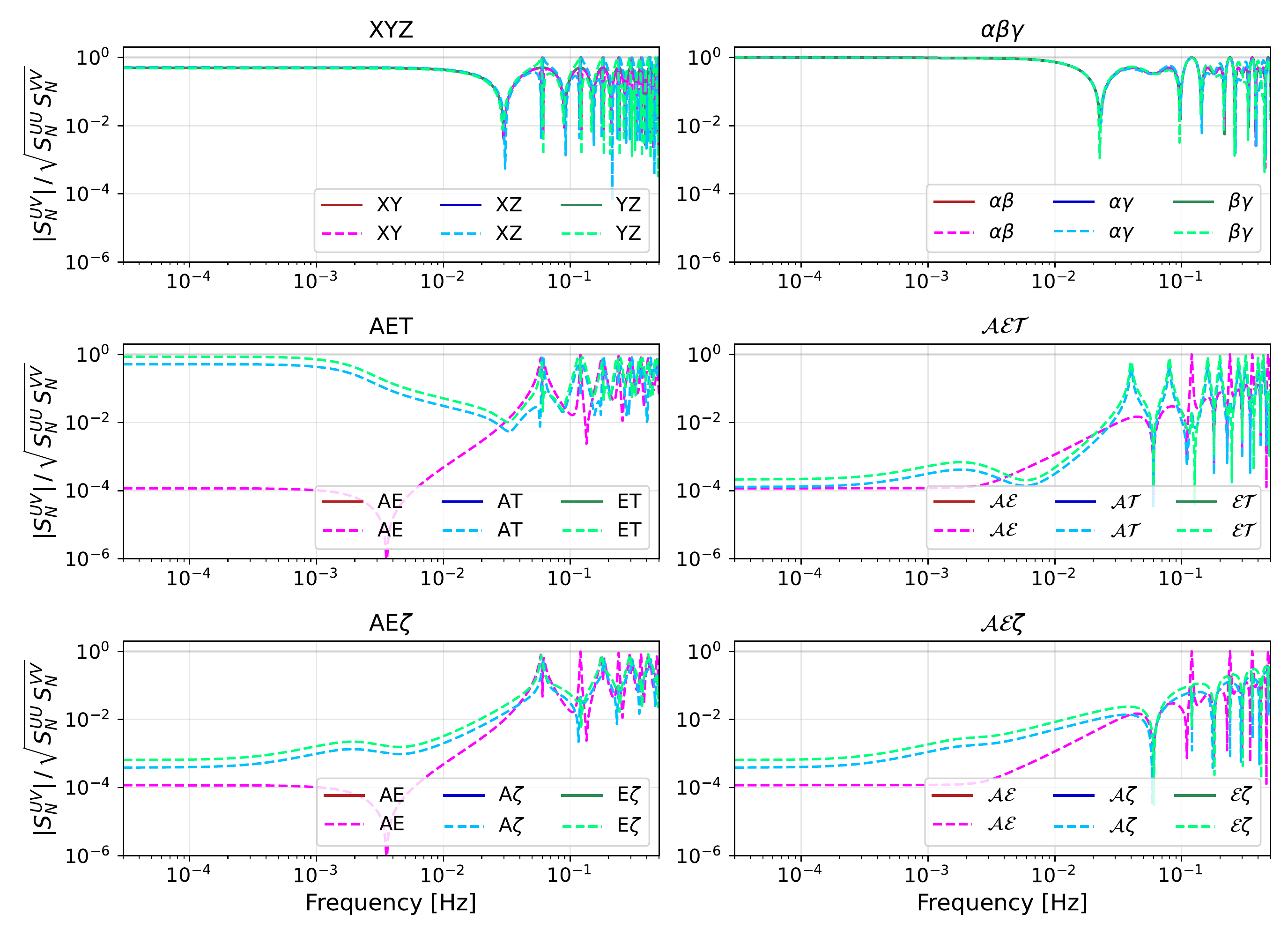}
\caption{Noise correlations considering equal noise levels for the TM acceleration and OMS noises and equal (solid) vs. unequal arms (dashed lines). The first column shows XYZ (first row), AET (second row), and AE$\zeta$. The second column shows $\alpha \beta \gamma$ (first row), $\mathcal{AET}$ (second row), and $\mathcal{AE}\zeta$.  }
\label{fig:noise_equal_correlations}
\end{figure}
In~\cref{fig:noise_equal_correlations} we show the square root of the noise coherence, defined by $|S^{UV, \mathrm{N}} | / \sqrt{ S^{UU, \mathrm{N}} \, S^{VV, \mathrm{N}}}$, for the three Michelson and Sagnac variables, for their orthogonal combinations, using the signal-insensitive variables T, $\mathcal{T}$ and $\zeta$. In the first row, we show the noise coherence for the base variables X, Y, and Z and $\alpha$, $\beta$, and $\gamma$. Both sets of three channels are strongly correlated, and the Sagnac channels are almost 100\% correlated at low frequencies.  In the second and third rows, we show the results obtained for the orthogonal channels. AE$\zeta$, $\mathcal{A}\mathcal{E}\mathcal{T}$ and $\mathcal{A}\mathcal{E}\zeta$ remain strongly uncorrelated (except close to the zeros). $\mathcal{A}\mathcal{E}\mathcal{T}$ is slightly less correlated than $\mathcal{A}\mathcal{E}\zeta$ at low frequencies while it is more strongly correlated in the oscillatory region at high frequencies.  AE$\zeta$ and $\mathcal{A}\mathcal{E}\zeta$ are very similar. T, however, becomes strongly correlated to A and E across a broad range of frequencies. Note that there are no solid lines in the second and third rows. Indeed, by construction, the orthogonal variables are fully uncorrelated when assuming equal arms and equal noise levels. \\
    
\begin{figure}[t]
\centering
\includegraphics[width=\linewidth]{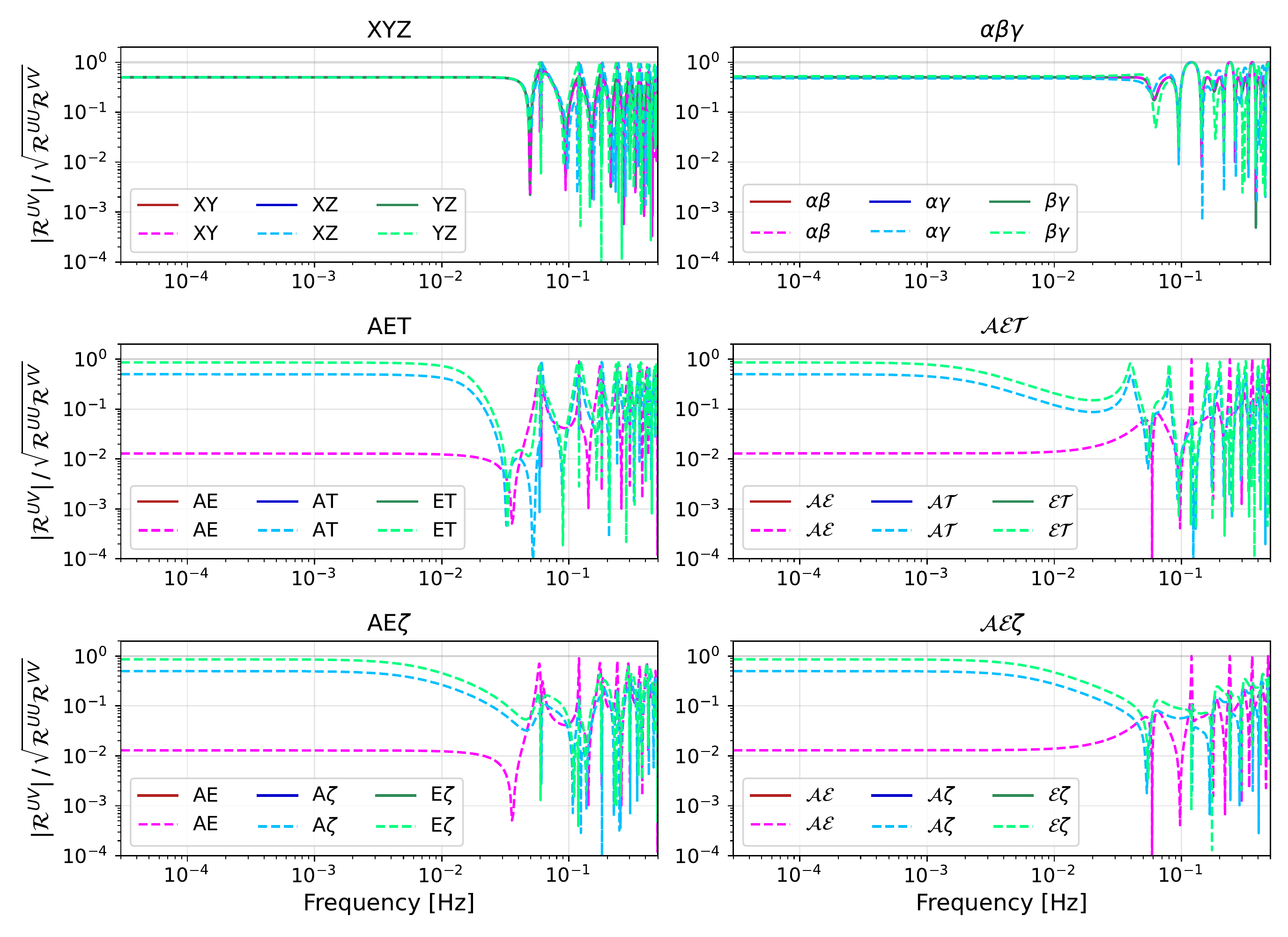}
\caption{Signal response correlations for the Michelson variables XYZ (first row, first column), the Sagnac variable $\alpha  \beta \gamma$ (first row, second column), their respective orthogonal channels AET (second row, first column), $\mathcal{A} \mathcal{E} \mathcal{T}$ (second row, second column) and combinations including the null channels AE$\zeta$ (last row, first column), $\mathcal{A} \mathcal{E} \zeta$ (last row, second column). Equal arms are shown with solid lines and unequal arms with dashed lines.}
\label{fig:signal_correlations}
\end{figure}

In~\cref{fig:signal_correlations} we show the square root of the signal response coherence $| \mathcal{R}^{UV} | / \sqrt{ \mathcal{R}^{UU} \mathcal{R}^{VV}}$ for all TDI variables considered. In the first row, we give the square root of the signal coherence for the base XYZ variables and for $\alpha \beta \gamma$. Both sets of TDI variables are strongly correlated. The equal and unequal arms models agree very well at low frequencies but show some differences near the peaks around the zeros at high frequencies. Note that while the noise spectra of $\alpha$, $\beta$, and $\gamma$ do not have any zeros in the LISA band, the CSDs do have zeros, but not as densely spaced as the ones for the Michelson variables. The second row shows the square root of the signal response coherence for the orthogonal AET and $\mathcal{A} \mathcal{E} \mathcal{T}$ variables. Note that A and E, and similarly $\mathcal{A}$ and $\mathcal{E}$, are not as correlated as the other pairs of TDI variables in the orthogonal sets. This is also true when T and $\mathcal{T}$ are replaced with $\zeta$, as shown in the third row, such that all null channels show strong low-frequency correlations with the other variables. We note that, at the same time, the T variable becomes significantly more sensitive to GWs than $\mathcal{T}$ or $\zeta$, as shown in in~\cref{fig:signal_psds_XYZ} and~\cref{fig:signal_psds_ABG}. This implies that the same loss in signal orthogonality for T, $\mathcal{T}$ and $\zeta$ more significantly impacts the overall orthogonality of AET than that of the other sets.

\subsection{Unequal noise levels}%
\label{sec:un_noise}
    
In this section, we consider the more general and realistic case in which all the individual TM and OMS noise levels, $A_{ij}$ and $P_{ij}$ (with $ij\in \{12,23,31,21,32,13\}$) are unequal. The values of $A_{ij}$ and $P_{ij}$ are generated as explained in~\cref{sec:noise_single_link} and, in the specific example we discuss here, they take the following values:
\begin{equation}
\label{eq:unequal_noise_levels}
A_{ij}  =  \{3.61, 3.02, 2.87, 3.43, 2.65, 3.45 \} \; , \qquad 
P_{ij}  =  \{ 14.00, 16.93, 9.43, 21.55, 17.04, 20.83 \}
\end{equation}
As before, the equal arms results are shown using solid lines while the unequal arms results are shown with dashed lines. Note that we do not show plots for the overall noise PSDs of the variables, as they remain qualitatively unchanged with respect to the equal noise case ones. In addition, given that only the instrumental noise levels are changed, the conclusions reached for the signal response and sensitivity discussed in~\cref{ssec:self-correlations-equal} remain valid.\\
\begin{figure}[htbp!]
\centering
\includegraphics[width=\linewidth]{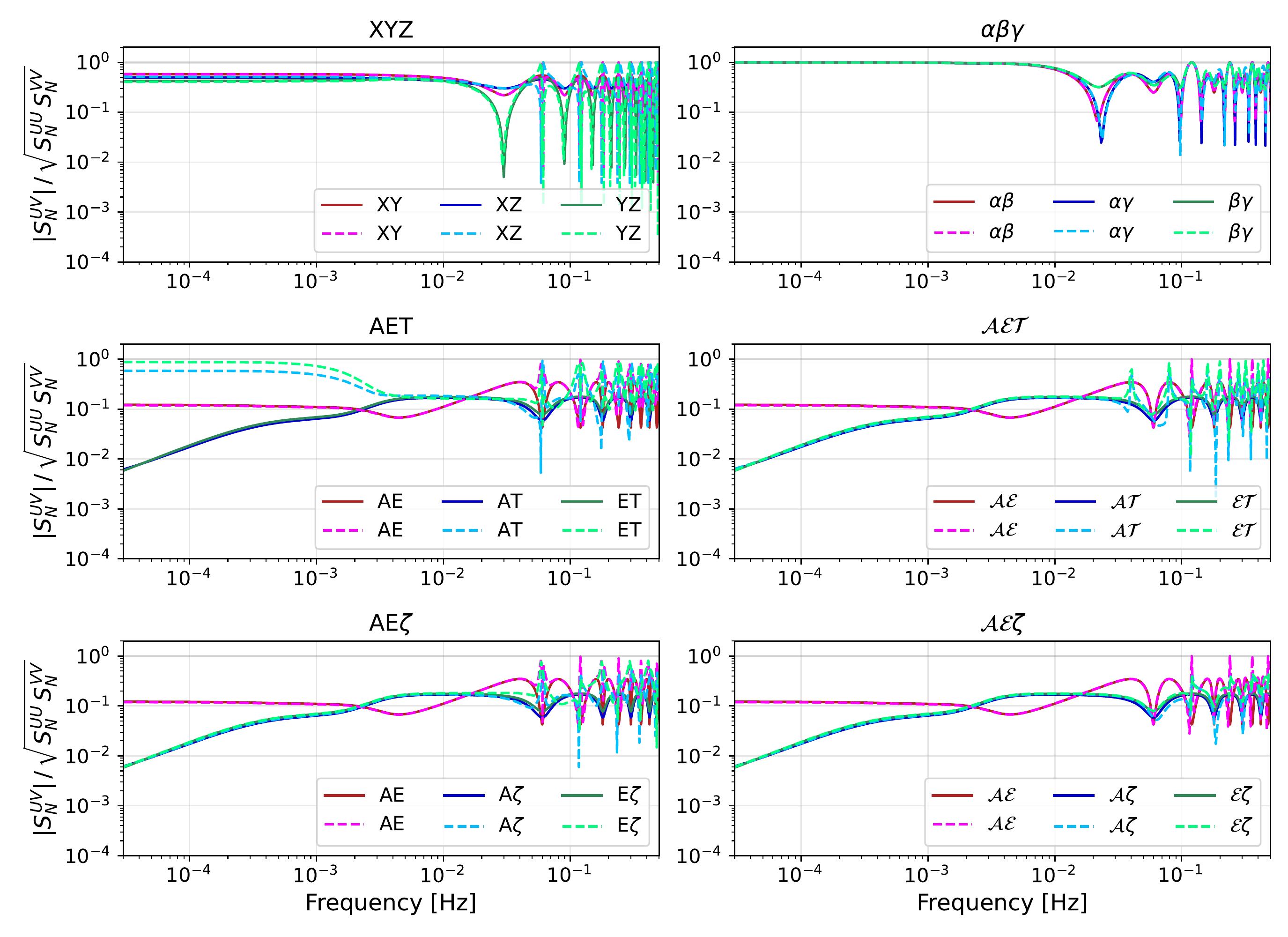}
\caption{Noise correlations considering six unequal levels for the TM and OMS noises and equal (solid) vs. unequal arms (dashed lines). The first column shows XYZ (first row), AET (second row), and AE$\zeta$. The second column shows $\alpha \beta \gamma$ (first row), $\mathcal{AET}$ (second row), and $\mathcal{AE}\zeta$.}
\label{fig:noise_unequal_correlations}
\end{figure}
 
The first row of~\cref{fig:noise_unequal_correlations} shows the square root of the noise coherence for X, Y, and Z and for $\alpha$, $\beta$, and $\gamma$. Both sets of variables behave similarly to the equal noise case, with the exception that some of the zeros are smoothed out and that the Michelson variables have slightly different levels of coherence at low frequencies. The results for the orthogonal channels are shown in the second row of the figure.  Both AET and $\mathcal{A}\mathcal{E}\mathcal{T}$ now show levels of coherence that reach $\sim 10\%$ for AE at all frequencies, and several percent for AT and ET even for equal arms. These levels are typically higher than those induced by the inequality of the arm-lengths discussed in the previous section, that is to say, the inequality in the levels of the noise has a stronger impact than the inequality of the LISA arms for most pairs of variables. The results obtained for unequal noise levels are approximately identical for all variables (with the exception of the peaks near the zeros), whether the LISA arms are equal or not, the only exception being T, which shows stronger correlations to A and E when the LISA arms are unequal. \\

Finally, the third row of~\cref{fig:noise_unequal_correlations} shows the correlations obtained when replacing T and $\mathcal{T}$ with $\zeta$. In both cases, the overall behaviour is very similar to that of $\mathcal{A}\mathcal{E}\mathcal{T}$, and seems to be dominated by the correlations due to unequal noise levels at almost all frequencies, such that the equal arm and unequal arm results are superimposed. Differences again appear near the peaks close to the zeros, for which $\mathcal{A}\mathcal{E}\zeta$ seems to slightly outperform both AE$\zeta$ and $\mathcal{A}\mathcal{E}\mathcal{T}$.\\
    
\section{SGWB and noise parameter reconstruction for AET and AE\texorpdfstring{$\zeta$}{zeta}}  %
\label{sec:parameter-reconstruction}
As shown in the previous section, relaxing the assumption that all arms are of equal length significantly breaks the orthogonality of the AET channels while that of AE$\zeta$, $\mathcal{A}\mathcal{E}\mathcal{T}$, and $\mathcal{A}\mathcal{E}\zeta$ is preserved to a large degree. Moreover, relaxing the assumption of equal secondary noises on each spacecraft also breaks the orthogonality of all TDI variable sets. In this section, we restrict ourselves to the unequal arms case and study how, in practice, neglecting the off-diagonal terms impacts the uncertainty in SGWB reconstruction in the equal and unequal noise cases. For each of the two cases, we perform Fisher forecasts, validated with Markov Chain Monte Carlo (MCMC) runs\footnote{The reader is referred to~\cref{sec:appendix_data_analysis} for the technical details of the MCMC data analysis.} to test the goodness of the Gaussian approximation. We restrict the following analysis to AET and AE$\zeta$.  The MCMC runs are performed considering only the diagonal terms of the AET and AE$\zeta$ TDI matrices\footnote{The main reason to restrict our analysis to this simplified scenario is that the present version of the data compression techniques described in~\cref{sec:appendix_data_analysis} are not suitable for application to problems including off-diagonal terms in the TDI matrix. Including those terms would thus require considering the full (and uncompressed) data set, which would significantly increase the computational cost of the MCMC runs.}. We proceed by comparing the results obtained with pure Fisher analyses, including, or neglecting, the off-diagonal terms in the TDI matrices, in order to assess their impact on the precision with which the determination of signal parameters can be made. A similar analysis for the noise parameters is presented in~\cref{sec:noise_analysis}. We conclude by comparing the MCMC posteriors for the signal parameters obtained with the different levels of complexity for the LISA configuration we have discussed throughout this work.\\
    
For all the analyses discussed in this section, we assume the LISA frequency range to be $3\times 10^{-5} \leq f\leq 5\times 10^{-1}$ Hz and the signal to be described by a simple power-law template:
\begin{equation}
\Omega_{\rm GW} h^2(f) = 10^{\alpha} \, \left( \frac{f}{f_*} \right)^{n_T} \; ,
\end{equation}
where the pivot frequency $f_*$ is chosen to be the geometric mean of the minimal and maximal frequency, i.e., $f_* \simeq 3.873 \times 10^{-3}$Hz. In all analyses presented in this section (and similarly in~\cref{sec:noise_analysis}), we consider a signal with zero slope, i.e., $n_{\rm T} = 0$, and $\alpha = -11.5$, corresponding to a Signal-to-Noise Ratio (SNR) $\simeq 271$ (see~\cref{eq:SNR_def}). Finally, The single link TDI transfer function is computed as outlined in~\cref{sec:signal_response} and the noise is modeled as indicated in~\cref{sec:noise_single_link}. Again, note that for brevity, we only focus on the AET and AE$\zeta$ TDI variables. \\

\begin{figure}[htbp!]
\centering
\includegraphics[width=\textwidth]{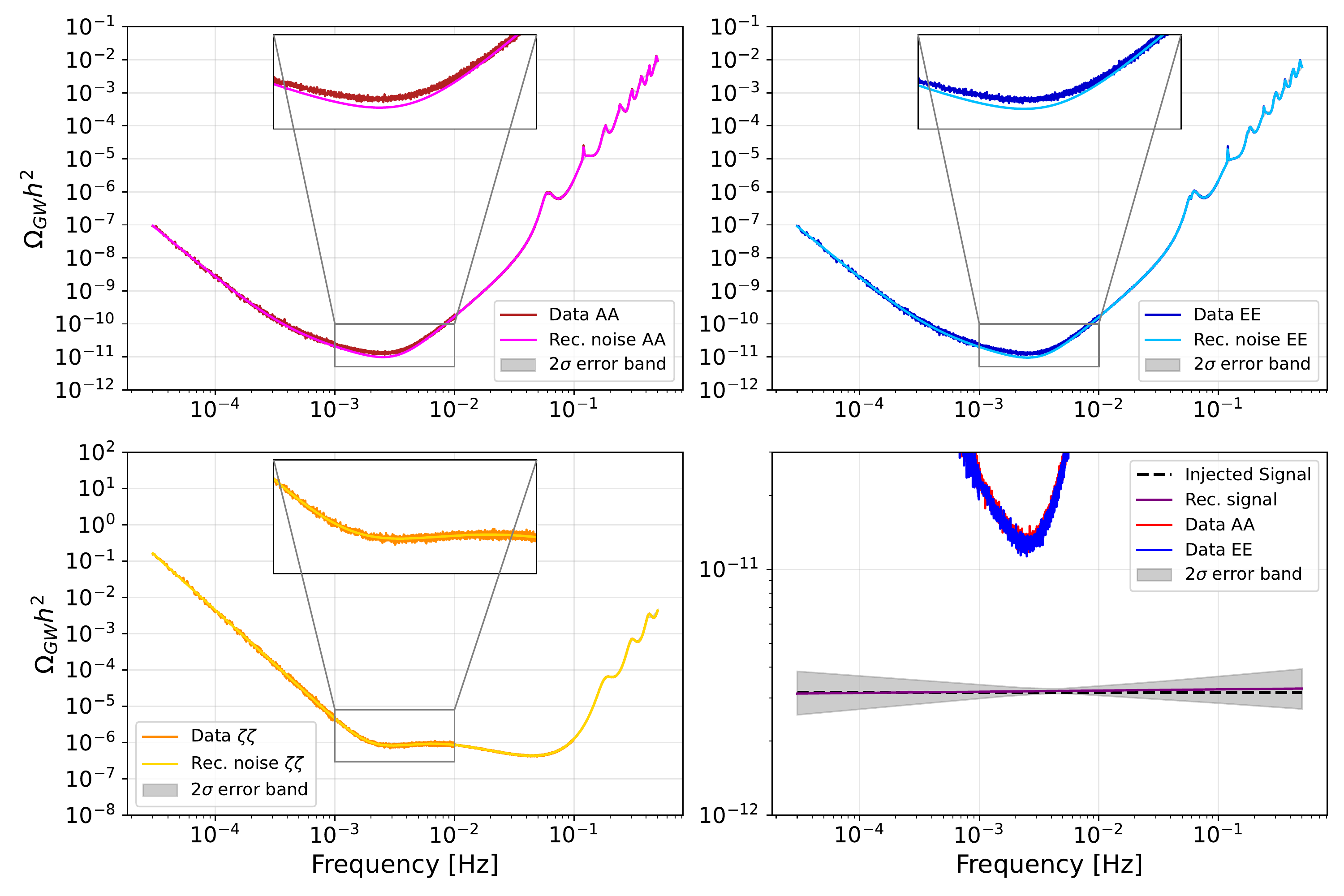}
\caption{Signal (bottom right) and noise (AA in top left, EE in top right, $\zeta\zeta$ in bottom left) reconstruction using the AE$\zeta$ TDI basis assuming unequal arm-lengths with equal noise levels.\label{fig:reconstruction_AEZ_eq}}
\end{figure}
    
In the following (and also in~\cref{sec:noise_analysis}), we denote quantities recentered on zero and normalized by the fiducial values with an overbar, and we denote quantities shifted with respect to the chain means and normalized using the fiducial value parameter values with a tilde.  Thus, while the latter provide information on the posterior widths but also on the best-fit parameter values, the former are best suited to compare posterior widths and shapes.\\
 
\begin{figure}[htbp!]
\centering
\includegraphics[width=\textwidth]{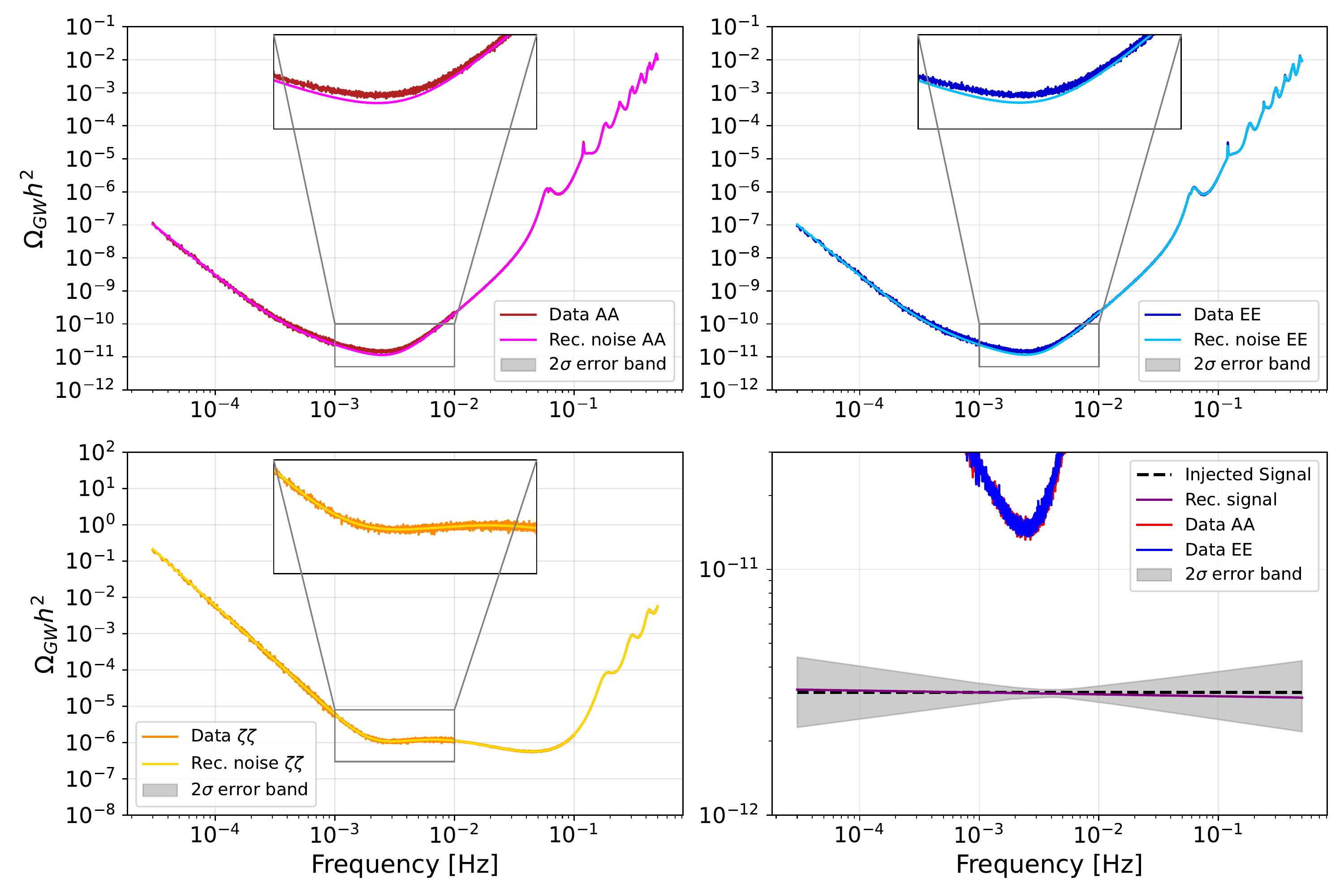}  
\caption{Signal (bottom right) and noise (AA in top left, EE in top right, $\zeta\zeta$ in bottom left) reconstruction using the AE$\zeta$ TDI basis assuming unequal arm-lengths with unequal noise levels.\label{fig:reconstruction_AEZ_un}}
\end{figure}

\begin{figure}[htbp!]
\centering
\begin{subfigure}{0.495\textwidth}
\includegraphics[height=5.5cm,keepaspectratio]{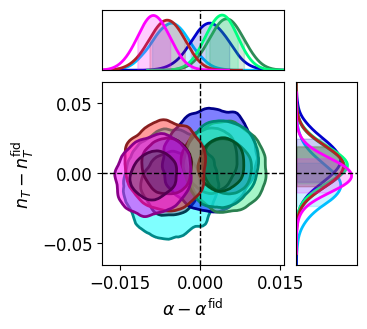}
\caption{SGWB parameters vs. fiducial. \label{fig:SGWB_fiducial}}
\end{subfigure}
\begin{subfigure}{0.495\textwidth}
\includegraphics[height=5.5cm,keepaspectratio]{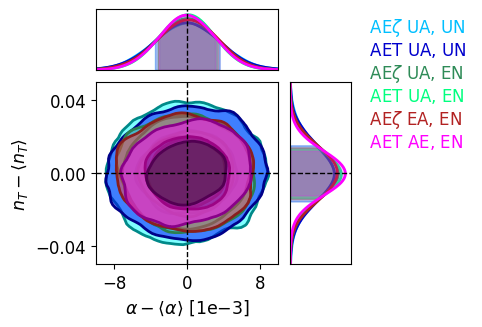}
\caption{SGWB parameters recentered on zero. \label{fig:SGWB_centered}}
\end{subfigure}
\caption{Comparison of MCMC SGWB parameter reconstruction for AET and AE$\zeta$, for all the configurations considered in this work, with UA/EA standing for un/equal arms, and UN/EN standing for un/equal noise amplitudes. Note that for this particular figure, the posteriors aren't normalized by the fiducial.  \label{fig:SGWB_contourplots}}
\end{figure}

Let us first present, in~\cref{fig:reconstruction_AEZ_eq}, the reconstruction of the signal and noise spectra (in $\Omega$ units) for unequal arms but equal noise levels using the diagonal of the AE$\zeta$ matrix. Both the signal parameters and the two noise parameters, and thus the signal and noise levels plotted in~\cref{fig:reconstruction_AEZ_eq}, are compatible at two sigmas with the injected signal and noise (as shown by the green contours in ~\cref{fig:SGWB_fiducial} for the signal and as shown also in~\cref{fig:allfreq2param}). Note that the noise reconstruction for the different TDI channels is so accurate that the error bands are not visible in~\cref{fig:reconstruction_AEZ_eq}.  On the other hand, while the reconstructed signal is still compatible with the injection, the error band for the signal, shown in the bottom right panel, is sufficiently large to be clearly visible. Although we do not include a plot of the reconstructed signal and noise spectra obtained for AET, the posteriors for the signal parameters are shown in~\cref{fig:SGWB_fiducial}, while those for the noise are shown in ~\cref{fig:allfreq2param}.\\

\begin{figure}[htbp!]
\centering
\begin{subfigure}{0.495\textwidth}
\includegraphics[height=5.5cm,keepaspectratio]{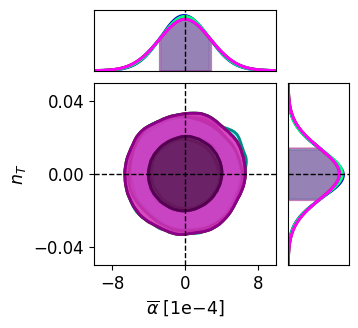}
\caption{Equal noise case.\label{fig:GWcontours_allfreq_2params}}
\end{subfigure}
\begin{subfigure}{0.495\textwidth}
\includegraphics[height=5.5cm,keepaspectratio]{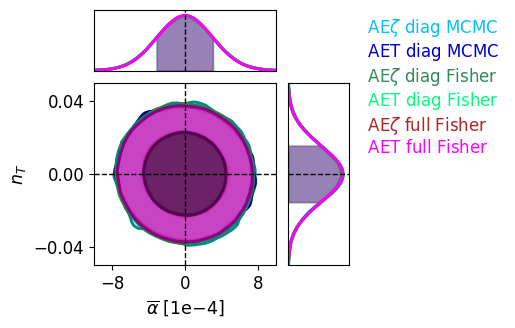}
\caption{Unequal noise case.\label{fig:GWcontours_allfreq_12params}}
\end{subfigure}
\caption{GW parameter posteriors in the case of unequal arms and {\it equal} or {\it unequal} noises. The results of the MCMC runs, the FIM analyses using the diagonal or the entire TDI covariance matrix are shown to superimpose for AET and $AE\zeta$. Note that the results are centered on zero and rescaled by the values of the fiducial parameters.\label{fig:GWcontours_allfreq}}
\end{figure}

Let us conclude our discussion of the unequal arms but equal noise case by comparing the results of the MCMC runs with those of the FIM analysis, see \cref{fig:GWcontours_allfreq_2params}.  This comparison focuses on the posterior widths and the degeneracies between parameters, such that all posteriors are recentered on zero.  This figure illustrates that there is nearly perfect agreement between the results obtained with the FIM analysis performed using the diagonal of the TDI covariance matrix and the MCMC results, and also between AE$\zeta$ and AET. It also shows a comparison with the FIM results that include the CSDs in the AET and AE$\zeta$ analyses and demonstrates that the reconstruction of the signal parameters would only be marginally affected by the inclusion of these terms.\\ 
    
Let us now discuss the case in which the TM and OMS noise levels are unequal, with amplitudes $A_{ij}$ and $P_{ij}$ given in~\cref{eq:unequal_noise_levels}. The signal and noise reconstructions in the AE$\zeta$ basis are shown in~\cref{fig:reconstruction_AEZ_un}. As for the equal noise case considered above, a detailed discussion of the noise parameter reconstruction is presented in~\cref{sec:appendix_unequal_noises}. Let us simply mention here that all the noise parameters are compatible with the injection parameters at the one or two-sigma level, see~\cref{fig:fullfreq12param_TM} and~\cref{fig:fullfreq12param_OMS}. As far as the signal reconstruction is concerned, we can see from the bottom-right panel of~\cref{fig:reconstruction_AEZ_un} and from the blue contours in ~\cref{fig:SGWB_fiducial}, that the reconstructed values are within the two sigma region from the injected values.  Once again, as can be seen in~\cref{fig:GWcontours_allfreq_12params}, we find excellent agreement between the FIM analysis and the MCMC results for both AET and AE$\zeta$, and find that the precision with which the signal parameters can be recovered is similar for both choices of TDI variables. Finally,  ~\cref{fig:GWcontours_allfreq_12params} also demonstrates that the results for the signal parameters remain the same whether one includes or excludes the off-diagonal terms of the TDI covariance matrix.\\

To sum up, using~\cref{fig:GWcontours_allfreq_2params} and~\cref{fig:GWcontours_allfreq_12params}, we can conclude that, for the simple power-law model considered in this work, the precision with which the reconstruction of the SGWB parameters can be achieved is not sensitive to the inclusion of the off-diagonal terms in the TDI covariance matrix, and that AET and AE$\zeta$ give comparable results.\\

Let us end this section by further commenting on the contour plots shown in~\cref{fig:SGWB_contourplots}. The figure provides a comparison of the MCMC results obtained for three different scenarios: 1) equal arms and equal noises, 2) unequal arms and equal noises, and lastly 3) unequal arms and unequal noises. \Cref{fig:SGWB_fiducial} demonstrates that the reconstructed signal is compatible with the injection parameters for all cases. \Cref{fig:SGWB_centered} demonstrates that the posterior widths remain largely unchanged across all scenarios.  These plots demonstrate that the reconstruction of the signal parameters is only marginally sensitive to the complexity of the underlying LISA scenario. This is one of the main results of the present work.\\

\section{Conclusions}%
 \label{sec:conclusion}

In this work, see in particular~\cref{sec:spectral-analysis},  we first studied the impact of a non-equilateral but stationary configuration for the LISA constellation and the effect induced by considering independent noise levels for each test mass (TM) and optical metrology system (OMS) on the orthogonality of the most well-known TDI bases, namely the Michelson XYZ, AET, and AE$\zeta$, and the Sagnac $\alpha\beta\gamma$, $\mathcal{A}\mathcal{E}\mathcal{T}$ and $\mathcal{A}\mathcal{E}\zeta$.  While the two noises propagate differently in the TDI variables, leading to different cross-correlations in the TDI bases, TM noise contributions dominate the low frequencies for signal-sensitive variables. On the other hand, for signal-orthogonal variables, OMS noise dominates the whole frequency range. For the case of equal noise levels, we confirmed the result, already known in the literature~\cite{Adams:2010vc}, that the null channel T, built from the Michelson XYZ variables, loses its signal and TM noise orthogonality if the constellation is not perfectly equilateral, becoming similar to the signal-sensitive variables. Interestingly, the loss of orthogonality between the signal-sensitive A and E channels and the null channel, T, is worse at low frequencies.\\

We also showed, for the first time, that other null channels, e.g., the $\mathcal{T}$ channel, built from the $\alpha\beta\gamma$ Sagnac TDI variables~\cite{Prince:2002hp}, or the variable $\zeta$~\cite{Armstrong_1999}, prove to be more robust under perturbations of the equilateral configuration. Moreover, these alternative TDI bases achieve the same level of laser noise suppression when extended to second generation, requiring fewer loops around the LISA satellites in their definitions~\cite{Muratore:2020mdf}. Accordingly, this would generate fewer zeros in the response functions, reducing the frequency window lost in the TDI definition process~\cite{Hartwig:2021mzw}. For {\it both} unequal arms and unequal noises, we demonstrated that {\it all} TDI bases exhibit sizeable cross-correlations in the noise and signal. For the T channel, the arm-length mismatch still produces the largest effect at low frequencies, while for the other channels, the impact of unequal noises is most significant across the whole band.\\

We then studied the signal and noise parameter reconstruction in~\cref{sec:parameter-reconstruction} and~\cref{sec:noise_analysis} respectively, using a Fisher Information Matrix (FIM) approach and Markov Chain Monte Carlo (MCMC) runs for two out of the total of six TDI bases, namely AET and AE$\zeta$. While the FIM analysis is a convenient tool that provides information on the posterior widths, the MCMC is computationally costly but also determines the best-fit parameters by confronting the signal and noise models with the LISA data. In our MCMC runs, for the reasons explained in ~\cref{sec:parameter-reconstruction}, we ignored the cross-correlations among various TDI variables. On the other hand, we used FIM analysis to compare results obtained when the cross-correlations were either included or excluded.\\

For the equal noise case, the large cross-correlations among the AET (quasi-)orthogonal TDI variables induced by the non-equilateral configuration of the LISA spacecraft result in a mild under-estimation of the TM acceleration amplitude parameter $A$ and a slight over-estimation of the OMS noise parameter $P$ in the signal and noise parameter reconstruction performed using MCMC and the FIM analysis. This can be understood using the toy model of~\cref{sub:toymodel}. Note that, as mentioned already, the frequency range over which the analysis is performed can play an important role, as shown by comparing the two contours plots of~\cref{fig:contours2param}. These findings do not apply to the $\mathcal{AE} \zeta$ TDI basis, which, rather than being sensitive to the arm-length mismatch, is only affected by the inequality in the noise amplitudes. Both in the equal and unequal noise cases, the OMS noise parameters are typically better determined than the TM noise parameters. This is mainly due to the fact that, while the information on the TM noises mostly comes from the low-frequency part of the frequency band, information on OMS noises comes from the high-frequency spectrum (for the null channels, OMS noise dominates the entire frequency range), which as larger weight in the likelihood since it contains more data points.  While the above discussion pertains to the MCMC runs, which were performed using only the diagonal of the TDI matrices, we remind the reader that, as shown in the FIM results of~\cref{fig:fullfreq12param_corr}, including cross-correlations has a significant impact on the uncertainty with which these noise parameters can be obtained.\\

Let us stress that even if some of the noise parameters are rather loosely constrained, the overall signal and noise shapes can be recovered with sufficient accuracy for all the scenarios tested in this work. While some previous analyses, e.g.,~\cite{Adams:2010vc, Adams:2013qma,Wang:2022sti}, have already considered different levels for the TM and OMS noises,  we also studied and quantified their impact on signal parameter reconstruction for different sets of TDI variables, going beyond the usual XYZ and AET bases. It is particularly noteworthy that the signal parameter reconstruction does not vary significantly among the different configurations considered, see~\cref{fig:SGWB_contourplots}, and that it is not sensitive to the inclusion of the off-diagonal terms of the TDI matrix, see~\cref{fig:GWcontours_allfreq_2params} and~\cref{fig:GWcontours_allfreq_12params}. For all these reasons, the results presented in this work represent an important contribution to the community's understanding of cosmological SGWB data analysis for LISA. \\

 We conclude by commenting on some of the assumptions of our work and on the possibility of relaxing them in future analyses. First, let us stress that a major limitation of our work resides in the fact that we assume perfect knowledge of the functional form of the instrumental noise, while it is possible that unknown sources of noise will be present in the real data.  Future studies are needed in order to test the impact of these effects on noise and signal reconstruction. For existing works setting the path in this direction, see, e.g.,~\cite{Karnesis:2019mph,Muratore:2022nbh, Baghi:2023qnq}.\\ 
 
 Let us point out that the present analysis does not include any of the possible time dependencies that might be present in the data. Beyond transients, the arm-lengths are expected to change over time, and the noise levels, or even the signal (e.g., the SGWB due to CGBs is expected to feature an annual modulation~\cite{Adams:2013qma}), might have some modulations over time, making the configuration non-static. Similarly, while we have restricted our study to isotropic SGWBs, the signal might have a non-trivial angular structure, whose reconstruction should be one of the targets to be included in an SGWB data analysis pipeline. Existing works in this direction range from more theoretical studies~\cite{LISACosmologyWorkingGroup:2022kbp} to numerical techniques involving the decomposition of the LISA response function in pixel space~\cite{Contaldi:2020rht} / in spherical harmonics~\cite{Banagiri:2021ovv}. \\
 
 Finally, in this work, we have considered a simplified scenario where a single SGWB of cosmological origin, described by a simple power-law mode with zero tilt, is present in the LISA band. In reality, as already mentioned in this work, at least two SGWB of astrophysical origin will be present, implying SGWB measurement will require component separation techniques to disentangle the different components contributing to the observed signal. See~\cite{Georgousi:2022uyt, Pozzoli:2023kxy} for studies estimating some of the astrophysical SGWBs for LISA, using an iterative source subtraction technique~\cite{Karnesis:2021tsh}, and see, e.g.,~\cite{Pieroni:2020rob, Flauger:2020qyi, Boileau:2020rpg, Boileau:2021gbr, Boileau:2022ter} for studies attempting a simultaneous detection of astrophysical and cosmological SGWBs. Furthermore, given that several cosmological mechanisms can generate spectra with a more elaborate frequency structure, one should test the robustness of the results obtained in this work when the assumption of a power-law signal is relaxed. Future studies aiming to be more realistic would have to include a combination of some of, and possibly all, these effects. 

\acknowledgements  
The authors thank Aurelien Hees and Peter Wolf, as well as Quentin Baghi, Jean-Baptiste Bayle, Valerie Domcke, Nikolaos Karnesis, Jonathan Gair, and Angelo Ricciardone for fruitful discussions. O.~H's work was supported by the Programme National GRAM of CNRS/INSU with INP and IN2P3 co-funded by CNES. O.H. and M.L. gratefully acknowledge support from the Centre National d'\'Etudes Spatiales (CNES). M.~M.~gratefully acknowledges support by the Deutsches Zentrum fur Luft- und Raumfahrt (DLR) with funding from the Bundesministerium fur Wirtschaft und Technologie (Project Ref. No. 50 OQ 2301 based on work done under Project Ref. No. 50 OQ 1801). M.~P.~'s work was partially funded by STFC grant ST/T000791/1 and by the European Union’s Horizon 2020 Research Council grant 724659 Massive-Cosmo ERC-2016-COG. M.~P.~thanks Imperial College London for hosting him during the early stages of this work. 

\appendix%
\FloatBarrier
\section{Useful relationships amongst TDI variables}%
\label{ssec:tdi-relations}
In this Appendix, we provide some useful relationships between the different TDI channels introduced in~\cref{sec:TDI_variables}. In particular, in~\cref{ssec:tdi-relations-uneqarm}, we show how the Sagnac variables are related to the Michelson variables in the unequal arms case, such that we would expect any independent set of three TDI channels to contain almost the same information. Then, in~\cref{ssec:tdi-relations-eqarm}, we give explicit relationships between the different sets of (quasi-)orthogonal channels. These expressions are given assuming equal LISA arms, where they can be formulated concisely as properties of the TDI variables themselves, regardless of the noise or signal correlations in the actual data. We also provide simplified low-frequency expansion of these relationships, which remain valid in the more general unequal-arm case for all covariance matrices not involving the Michelson T channel.

\subsection{Unequal arms\label{ssec:tdi-relations-uneqarm}}%

It is known that for a constellation with three constant, but unequal arms, one can exactly reproduce any TDI variable as a linear combination of four generators, with the set $\{\alpha,\beta, \gamma, \zeta\}$ as one possible basis~\cite{Dhurandhar:2001tct}. Furthermore, these four generators are themselves not fully independent, but can be related by~\cite{Armstrong_1999}:
\begin{equation}
(1 - \dx \dy \dz )\zeta = (\dx - \dy\dz)\alpha + (\dy - \dx \dz)\beta + (\dz - \dx \dy)\gamma\, . 
\end{equation}
This means that we can derive time-delay relationships between different variables using just three variables as a basis. For example, the following relationships can be used to express $\alpha$ and $\zeta$ in terms of X, Y, and Z:
\begin{subequations}
\begin{align}
\begin{split}
(\dx^2-1)(\dy^2-1)(\dz^2-1) \alpha &= (1 + \dx\dy\dz)(\dx^2 - 1) X \\
&\quad+ (\dx\dy + \dz)(\dy^2  -1 ) Y \\
&\quad + (\dy + \dx\dz)(\dz^2 - 1) Z 
\end{split}
\\
\begin{split}
(\dx^2-1)(\dy^2-1)(\dz^2-1) \zeta &= (\dx + \dy\dz)(1 - \dx^2) X \\
&\quad+ (\dy + \dx\dz)(1 - \dy^2) Y \\ 
&\quad+ (\dz + \dx\dy)(1 - \dz^2) Z \; .
\end{split}
\end{align}    
\end{subequations}
If we exclude Fourier frequencies at which some of the delay operator combinations in the previous equations lead to an exact cancellation of the signal\footnote{For instance, the Fourier transform of terms of the type $(1 - D_{ij}^N)$ is $(1 - e^{-2 \pi f N d})$, which is exactly zero if $f d$ is an integer. At such ``singular'' frequencies, the response of one variable can be exactly zero (while that of another is not). In this case, the two variables are no longer equivalent. In reality, we expect these zeros to be smoothed out to some extent due to numerical limitations and other noise sources, such that the discrepancies between two related variables will be extended to a small frequency band around each zero~\cite{Vallisneri:2005ji,Hartwig:2021mzw}.}, we expect all sets of three independent TDI variables to contain exactly the same information. This is verified by the fact that we get identical results in all cases when using the full TDI covariance matrices.\\

However, as we discuss in this paper, some TDI variables, such as the traditionally used Michelson T channel, prove to be particularly sensitive to deviations from the equal-arm assumption. This means they potentially require more elaborate models to achieve the same scientific output as more robust channels, such as $\mathcal{T}$ or $\zeta$. Furthermore, the exact cancellation of the TDI channels at some Fourier frequencies leads to zeros in the corresponding covariance matrix, making it non-invertible. Since parameter estimation typically requires inverting the noise covariance matrix this favors variables with fewer zeros, for which fewer Fourier frequencies need to be excluded from the analysis. $\mathcal{A}\mathcal{E}\zeta$ is the optimal set considered in this work from this viewpoint, with the first singularity of the covariance matrix appearing at $f = 1/L \approx \SI{0.12}{\hertz}$.
    
\subsection{Relationship between orthogonal channels in the equal arms limit\label{ssec:tdi-relations-eqarm}}%
In the limit of equal LISA arms, we can derive simple relationships between the TDI coefficient matrices $C^{UV}$, as defined in~\cref{eq:TDI-cross-spectra}, for the different sets of quasi-orthogonal channels considered in this manuscript. We can further perform a low-frequency expansion of these relationships, the results of which are given alongside the full expressions after the $\simeq$ signs:
\begin{subequations} \label{eq:orthogonal-relationships-equalarms}
\begin{alignat}{3}
&C^{ \rm{AA} / \rm{EE} / \rm{AE}} = 4 \cos^2(\pi f L)C^{ \mathcal{A} \mathcal{A} / \mathcal{E}\mathcal{E} / \mathcal{A}\mathcal{E} } \qquad &\simeq &\, 4 C^{ \mathcal{A} \mathcal{A} / \mathcal{E}\mathcal{E} / \mathcal{A}\mathcal{E} }\,, \\
&C^{ \rm TT } = \frac{16}{3} \sin ^2\left(\pi f L \right) \sin ^2\left(2 \pi f L \right) C^{\zeta\zeta} \qquad &\simeq &\, \frac{64}{3} L^4 \pi^4 f^4 C^{\zeta\zeta}\,,\\
&C^{ \mathcal{T} \mathcal{T}  } = \frac{\left[ 1 + 2 \cos(2 \pi f L ) \right]^2}{3} \; C^{\zeta\zeta} \qquad &\simeq &\, 3 C^{\zeta\zeta}\,,\\
&C^{ \rm{AT} / \rm{ET}} = -\frac{4\sin^2(2 \pi f L )}{1+2\cos(2 \pi f L )} C^{ \mathcal{A} \mathcal{T} / \mathcal{E} \mathcal{T} }  \qquad &\simeq &\,  - \frac{16 L^2 \pi^2 f^2}{3} C^{ \mathcal{A} \mathcal{T} / \mathcal{E} \mathcal{T} }\,,\\
&C^{ \rm{AT} / \rm{ET}} = \frac{e^{4 i \pi f L } - 2 i \sin(2 \pi f L ) - 1}{\sqrt{3}} C^{ {\rm A} \zeta / {\rm E} \zeta}  \qquad &\simeq &\,  - \frac{8 L^2 \pi^2 f^2}{\sqrt{3}}  C^{{\rm A} \zeta / {\rm E} \zeta}\, , \\
&C^{\mathcal{A}\mathcal{T} / \mathcal{E}\mathcal{T}} = \frac{1 + 2 \cos (2 \pi f L )}{\sqrt{3}} C^{\mathcal{A}\zeta / \mathcal{E}\zeta} \qquad &\simeq &\, \sqrt{3} C^{\mathcal{A}\zeta / \mathcal{E}\zeta}\, .
\end{alignat}
\end{subequations}
The notation $C^{ UV / WL }$ indicates that the same expression is valid for $C^{UV}$ and $C^{WL}$ independently, but does not imply any relationship between $C^{UV}$ and $C^{WL}$.\\
    
Let us first note that the coefficient matrices of the different sets of quasi-orthogonal channels can be related by overall frequency-dependent scaling factors. This implies that the CSD and PSD terms in the matrix $S^{UV}$ (see~\cref{eq:TDI-cross-spectra}) will inherit the same relationships, irrespective of the single-link correlation matrix $S^\eta$. Since this applies equally to either noise or signal in the data, this further implies that the different sets all have almost exactly the same signal-to-noise ratio and are therefore almost equivalent for the purpose of data analysis, at least in the equal-arm approximation. The only caveat of this statement, as mentioned already in \cref{ssec:tdi-relations-uneqarm}, is that the frequency-dependent factors can be vanishing at singular frequencies. Therefore, the ``simpler'' variables with fewer zeros, such as $\mathcal{A}$, $\mathcal{E}$ or $\zeta$, in principle contain slightly more information than the more ``complex'' ones, like A, E, T, and $\mathcal{T}$. \\

Going to the low-frequency expansion simply gives a constant scaling factor for most cases. The T channel, however, shows a significantly stronger low-frequency suppression when compared to the otherwise equivalent $\mathcal{T}$ or $\zeta$ channels. We remark that T is also the only channel for which these low-frequency expansions do not remain valid approximations for the unequal arm scenario. This might explain why we find it to be more susceptible to deviations from the equal-arm assumptions at low frequencies.
    
\section{Noise and signal analytic approximations\label{sec:analytical-models}}%
In this appendix, we provide the analytic expressions of the noise PSDs and CSDs for the different configurations considered in this work.  In~\cref{sec:analytical-models}, we consider identical TM and OMS noise terms on all spacecraft for both the Michelson and Sagnac variables, as well as for their orthogonal channels, for the case of equal LISA arms. In~\cref{sub: un_noise psd}, we provide the corresponding expressions in the case of unequal but constant arms. Given that the full expressions are not particularly enlightening, we shall write down only their low-frequency (i.e., $f \ll c / (2 \pi L )$) expansions. The full expressions are made available as supplementary material. It is possible to compute the signal response over the entire LISA band by performing a numerical integration of ~\cref{eq:Upsilon}. In order to obtain an analytic expression, one instead has to expand this expression at low frequencies. We do so for equal and unequal arms in~\cref{sec:signal_resp_low_f} and~\cref{sec:signal_resp_low_f_unequal}, respectively.\\

All the expressions involving unequal arms which we consider below are computed by assuming that LISA can undergo two main distortion modes, namely $\delta_c$ and $\delta_d$~\cite{Muratore:2021uqj}. These distortions leave the average arm-length unchanged and can therefore be used to characterize arm-length mismatches across the three LISA arms. This approach helps in identifying the effects of each mode on the PSDs and CSDs of the TDI variables.
    
\subsection{Noise PSDs and CSDs for an equal arms configuration}%
\label{sub: noise psd}
Let us write down the expressions for the noise CSDs and PSDs of all channels assuming equal arms. We have
\begin{subequations}
\begin{align}\label{eq:michelson-noise-equal-arm}
S^{ \rm XX,N} = S^{ \rm YY, N} = S^{ \rm ZZ, N} &= 16 \sin ^2( 2 \pi f L ) \left\{  \left[ 3  + \cos (4 \pi f L) \right] S^\text{TM} + S^\text{OMS}  \right\} \; , \\ 
S^{ \rm XY, N} = S^{ \rm XZ, N} = S^{ \rm YZ, N} &= -4 \sin ( 2 \pi f L ) \sin (4 \pi f L ) \left( S^\text{OMS} +4 S^\text{TM} \right) \; , 
\end{align}
\end{subequations}
for the Michelson variables, and
\begin{subequations}
\begin{align}
S^{\alpha\alpha, \rm N} = S^{\beta\beta, \mathrm{N}} = S^{\gamma\gamma, \mathrm{N}} &= 6 S^\text{OMS} + 4 \left[3 - 2 \cos ( 2 \pi f L ) -\cos (6 \pi f L ) \right] S^\text{TM} \; ,  \\ 
S^{\alpha\beta, \mathrm{N}} = S^{\alpha\gamma, \mathrm{N}} = S^{\beta\gamma, \mathrm{N}} &= 2 \left[  2 \cos ( 2 \pi f L )+\cos (4 \pi f L) \right] S^\text{OMS} - 4 \left[ 1 - \cos ( 2 \pi f L ) \right] S^\text{TM} \; , 
\end{align}
\end{subequations}
for the Sagnac variables.\\
    
For the orthogonal channels A, E, we find:
\begin{subequations}
\begin{align}
S^{ \rm AA, N} = S^{ \rm EE, N} & = 8 \sin ^2( 2 \pi f L ) \left\{ S^\text{OMS} \left[ \cos ( 2 \pi f L )+2 \right]+2  \left[  3 + 2 \cos ( 2 \pi f L)+\cos (4 \pi f L ) \right] S^\text{TM} \right\} , \\
S^{ \rm AE, N} = S^{ \rm AT, N} = S^{ \rm ET, N} &= 0 \; .
\end{align}
\end{subequations}
As expected, the CSDs between the orthogonal channels in the case of equal arm-lengths are zero, since A, E, and T are defined to be orthogonal. Finally, the expression for $\zeta$ reads:
\begin{equation}
S^{\zeta\zeta, \mathrm{N}} = 6 \left\{ S^\text{OMS} +2 \left[ 1 - \cos ( 2 \pi f L ) \right] S^\text{TM} \right\} \; .
\end{equation}
The expressions for any other set of orthogonal variables follow from~\cref{eq:orthogonal-relationships-equalarms}.
    
\subsection{Noise PSDs and CSDs for an unequal arm-lengths configuration} \label{sub: un_noise psd}%
While it is possible to obtain exact expressions for the noise CSDs and PSDs in the unequal arm case, as mentioned before, the resulting expressions are rather large and cumbersome, and thus not particularly enlightening. For this reason, we shall only write down simpler expressions, by working in the low-frequency limit. Furthermore, as discussed in~\cite{Muratore:2020mdf}, one can express the arm-lengths $L_{ij}$ in terms of the breathing modes of the LISA triangle, $\delta_c$ and $\delta_d$~\cite{nussbaum1968} as:
\begin{subequations}\label{Lterm2}
\begin{align}
L_{23}(t)&= L \left[1 + \frac{1}{2}\left(\sqrt{3}\,\delta_c - \delta_d\right)\right] \; ,\\
L_{31}(t)& = L \left(1 + \delta_d\right) \; ,\\
L_{12}(t) &= L \left[1 - \frac{1}{2}\left(\sqrt{3}\,\delta_c + \delta_d\right)\right] \; .
\end{align}
\end{subequations}
and further expand the noise CSDs and PSDs in powers of $\delta_c$ and $\delta_d$.  Indeed, while $L = \frac{L_{12} + L_{23} + L_{31}}{3}\approx \SI{8.3}{\second}$ is the average arm-length, the small parameters $\delta_c $ and $\delta_d$ are typically of the order \SI{1}{\milli\second} to \SI{10}{\milli\second} for realistic orbits. Evidently, the case $\delta_c = \delta_d = 0$ corresponds to the equal LISA arms scenario.\\
    
Aside from $S^{\rm TT, N}$, none of the PSDs are modified by the arm-length mismatch at leading order in $\delta_{c,d}$.  In particular, we have\footnote{Expanding the TDI coefficient matrix $C^{TT}$ computed according to~\cref{eq:TDI-cross-spectra} to leading order in frequency shows that it \emph{only} contains terms that are second order in 
 $\delta_{c}$ and $\delta_d$. This implies that T's leading order dependence on the arm-length mismatch is a generic effect, and we should expect $S^{\rm TT, N}$ to exhibit low-frequency deviations for any kind of noise correlations we model in $S^\eta$.}:
\begin{equation}
S^{ \rm TT, N} \simeq 12 (\delta_c^2 + \delta_d^2) (2 \pi f)^2 L^2  (S^\text{OMS} + 4 S^\text{TM}) \;   . 
\end{equation}
Furthermore, none of the CSDs between quasi-orthogonal channels remain exactly zero if the arm-lengths are unequal:
\begin{subequations}
\begin{align}
S^{ \rm AE, N} &= -12 \delta_c \delta_d L^2 (2 \pi f)^2 (S^\text{OMS}+4 S^\text{TM}) = 4 S^{\mathcal{A}\mathcal{E}, \rm N} \; , \\
S^{ \rm AT, N} &= 12 \sqrt{2} \delta_c L^2 (2 \pi f)^2 (S^\text{OMS}+4 S^\text{TM}), \\
S^{ \rm A\zeta, N} &= \sqrt{6} \delta_c L^2 (2 \pi f)^2 (S^\text{OMS}+12 S^\text{TM}) \; , 
\\
S^{ \rm ET, N} &= 12 \sqrt{2} \delta_d L^2 (2 \pi f)^2 (S^\text{OMS}+4 S^\text{TM}),
\\
S^{ \rm E\zeta, N} &= \sqrt{6} \delta_d L^2 (2 \pi f)^2 (S^\text{OMS}+12 S^\text{TM}) \; , 
\\
S^{ \mathcal{A}\mathcal{T}, \rm N} &= - 6 \sqrt{2} \delta_c L^2 (2 \pi f)^2  S^\text{TM},
\\
S^{ \mathcal{A}\zeta, \rm N} &= 2 \sqrt{6} \delta_c L^2 (2 \pi f)^2 (S^\text{OMS}+3 S^\text{TM}) \; , 
\\
S^{ \mathcal{E}\mathcal{T}, \rm N} &= - 6 \sqrt{2} \delta_d L^2 (2 \pi f)^2  S^\text{TM},
\\
S^{ \mathcal{E}\zeta, \rm N} &= 2 \sqrt{6} \delta_d L^2 (2 \pi f)^2 (S^\text{OMS}+3 S^\text{TM}) \; .
\end{align}
\end{subequations}
We note that $S^{\rm AT}$ and $S^{ \rm ET}$ are proportional to $\delta_c$ and $\delta_d$, and are thus 2 orders of magnitude larger than $S^{ \rm TT}$, assuming $\delta_c \approx 0.01$.  This highlights the importance of the CSDs in the low-frequency regime when using A, E, and T with unequal arms. This is shown explicitely in~\cref{sec:appendix_equal_noises} in a comparison of $AET$ and $AE\zeta$ using FIM and MCMC in the low frequency range. $S^{\rm AE}$, on the other hand, is proportional to $\delta_c \delta_d$, and is therefore 4 orders of magnitude smaller than $S^{\rm AA}$ or $S^{ \rm EE}$. $S^{A\zeta}$ and $S^{E\zeta}$ are again proportional to $\delta_c$ and $\delta_d$, such that they are also suppressed by two orders of magnitude with respect to the diagonal terms $S^{\rm AA}$, $S^{ \rm EE}$ and $S^{\zeta\zeta}$. We can therefore conclude that A, E, and $\zeta$ remain almost orthogonal even in the unequal arms case. This is consistent with the discussion of~\cref{sec:spectral-analysis} and with the results of~\cref{sec:appendix_equal_noises}.
    
\subsection{Low-frequency signal response for an equal arm-length configuration}%
\label{sec:signal_resp_low_f}
    
One can obtain an analytic expression for~\cref{eq:Upsilon} by first expanding its kernel for low frequencies.  Doing so, one obtains in turn an analytic expression for the $S^{\eta,\mathrm{GW}}_{ij,mn}(f)$ CSD given by~\cref{eq:CSDsignal}.  This expression can then be used in~\cref{eq:TDI-cross-spectra} with the TDI coefficients, and the resulting expression expanded in powers of frequency. By doing so, one obtains the following expressions in the equal arm approximation:
\begin{subequations}
\begin{align}\label{eq:michelson-signal-equal-arm}
S^{ \rm XX,GW} = S^{ \rm YY, GW} = S^{ \rm ZZ, GW} &= \frac{384}{5} (f L \pi)^4 \; , \\ 
S^{ \rm XY, GW} = S^{ \rm XZ, GW} = S^{ \rm YZ, GW} &= - \frac{1}{2} S^{ \rm XX,GW} \; ,
\end{align}
\end{subequations}
for the Michelson variables, and:
\begin{subequations}
\begin{align}
S^{\alpha\alpha, \rm GW} = S^{\beta\beta, \rm GW} = S^{\gamma\gamma, \rm GW} &= \frac{96}{5} (f L \pi)^4 \; ,  \\ 
S^{\alpha\beta,\rm GW} = S^{\alpha\gamma,\rm GW} = S^{\beta\gamma,\rm GW} &= -\frac{1}{2} S^{ \alpha\alpha,GW} \; , \end{align}
\end{subequations}
for the Sagnac variables. Finally, we can compute equivalent expressions for AET, AE$\zeta$, $\mathcal{A}\mathcal{E}\mathcal{T}$ and $\mathcal{A}\mathcal{E}\zeta$:
\begin{subequations}
\begin{align}
S^{ \rm AA, GW} &= S^{ \rm EE, GW}  = \frac{3}{2} S^{ \rm XX,GW} \; ,  \qquad 
S^{ \mathcal{A}\mathcal{A},\rm GW} = S^{ \mathcal{E}\mathcal{E},\rm GW}  = \frac{3}{2} S^{ \alpha\alpha,\rm GW} \; , 
\\ \label{eq:null-channel-equal-arm-low-freq}
S^{ \rm TT,\rm GW} &= \frac{256}{63} (f L \pi)^{10},\qquad S^{ \mathcal{T}\mathcal{T},\rm GW} = \frac{4}{7} (f L \pi)^6 \; , \qquad
S^{ \zeta\zeta,\rm GW} = \frac{4}{21} (f L \pi)^6 \; ,
\end{align}
\end{subequations}
where all cross-terms are vanishing.
\subsection{Low-frequency signal response for an unequal arm-length configuration}%
\label{sec:signal_resp_low_f_unequal}
In order to obtain reasonably short expressions in the case of unequal arms, we follow the same procedure as in~\cref{sec:signal_resp_low_f} but also expand the resulting expressions in powers of the breathing modes $\delta_c$ and $\delta_d$ introduced in~\cref{Lterm2}. 
Similarly to the results obtained in~\cref{sub: un_noise psd}, we again find that most channels are unaffected by the leading order in $\delta_c$ and $\delta_d$. Contrary to~\cref{sub: un_noise psd}, we now find that all three null channels T, $\mathcal{T}$, and $\zeta$ are modified with respect to the equal arm case. Concretely, we get
\begin{equation}
S^{ \rm TT, GW} = 16 S^{ \mathcal{T} \mathcal{T}, GW} = \frac{16}{3}S^{ \zeta \zeta, GW} = \frac{288}{5}f^4\pi^4L^4\left(\delta_c^2+\delta_d^2\right)\;,
\end{equation}
such that each of the three channels now shows the same low-frequency behaviour as the GW-sensitive channels. However, comparison with~\cref{eq:null-channel-equal-arm-low-freq} shows that for $\mathcal{T}$ and $\zeta$, this deviation corresponds to a change in slope by $f^2$ with respect to the equal arm scenario, while for T, the slope changes by a factor $f^6$. This, together with the findings in~\cref{sub: un_noise psd}, explains the qualitatively very different behaviour between the T channel and the other null channels observed in~\cref{sec:spectral-analysis}. Furthermore, we observe that we now have $S^{ \mathcal{T} \mathcal{T}, GW} = \frac{1}{3} S^{ \zeta \zeta, GW}$, whereas we had $S^{ \mathcal{T} \mathcal{T}, GW} = 3 S^{ \zeta \zeta, GW}$ in the equal arm case. Since the noise curves of these two channels are, at leading order, unaffected by the arm-length mismatch, which explains why we found $\mathcal{T}$ to be a slightly better null channel than $\zeta$ in~\cref{sec:spectral-analysis}. \\

Finally, we note that while the CSDs of the base Michelson XYZ and Sagnac $\alpha\beta\gamma$ variables are again unaffected at leading order, all orthogonal channels exhibit non-vanishing cross-terms, given by:
\begin{subequations}
\begin{align}
S^{ \rm AE, GW} &= -\frac{\sqrt{2}}{2}\, S^{ \rm AT, GW} = -\frac{576}{5}f^4 \pi^4 L^4 \delta_c \;,\\
S^{ \rm ET, GW} &= \frac{288}{5}\sqrt{2}f^4\pi^4L^4\delta_d\;,\\
S^{\rm{A}\zeta,\rm GW} &= \frac{72}{5}\sqrt{6}f^4\pi^4L^4\delta_c\;,\\
S^{\rm{E}\zeta,\rm GW} &= \frac{72}{5}\sqrt{6}f^4\pi^4L^4\delta_d\;,\\
S^{ \mathcal{A} \zeta, GW} &= \frac{\sqrt{6}}{4}S^{ \mathcal{A} \mathcal{E}, GW} = -\frac{36}{5}\sqrt{6}f^4\pi^4L^4 \delta_c\;,\\
S^{ \mathcal{A} \mathcal{T}, GW} &= -\frac{36}{5}\sqrt{2}f^4\pi^4L^4\delta_c\;,\\
S^{ \mathcal{E} \zeta, GW} &= \sqrt{3}S^{ \mathcal{E} \mathcal{T}, GW} = -\frac{36}{5}\sqrt{6}f^4\pi^4L^4\delta_d\;.
\end{align}
\end{subequations}
    
\section{Data analysis technical details}%
\label{sec:appendix_data_analysis}
    
As discussed in~\cref{sec:measurement} a data stream of LISA (and in general of any GW experiment) can be modeled as a superposition of some signal and detector noise. Let us assume that the data $d_i(t)$, with the index $i$ labeling the TDI channel, are provided in the time domain, and for simplicity let us assume $d_i(t)$ to be stationary. This corresponds to assuming rigid arms during the full mission duration. This assumption is clearly not going to be valid for LISA, but for the sake of our discussion, which rather focuses on the impact of unequal arm-lengths and noise levels on the orthogonality, as well as on the constraining power, of the different TDI variables, we restrict ourselves to this simplified scenario. Given the total observation time of the detector $T_{d}$, which in this paper is always assumed to be $4$ years with $100\%$ efficiency, we can divide it into a given number, say $N_d$, of segments, of duration $T_{d}/N_d$ each. We can thus define the data $\tilde{d}^s_i(f_\textmd{k})$, in the frequency domain, where the index $s$ runs over all segments, and the index $\textmd{k}$ runs over frequencies in the detector range. The frequency resolution for any given segment is directly given by $\Delta f = N_d/T_{d}$. In the following, we will assume the time duration of each segment to be $\sim$ 11.5 days corresponding to $\Delta f \sim 10^{-6}$ Hz. We also assume different frequencies to be uncorrelated. As discussed in~\cref{sec:measurement}, both signal and noise are also assumed to be Gaussian-distributed with vanishing mean and variance given by their respective power spectral densities. Under all these assumptions, we can generate $N_d$ statistical realizations of the signal and of all the noise components. Following the procedure described in~\cite{Caprini:2019pxz, Flauger:2020qyi}, we define a new set of (averaged) data $\bar{D}^\textmd{k}_{IJ} \equiv \tilde{d}^i_J(f_\textmd{k}) \tilde{d}^i_I(f_\textmd{k}) / N_d$, which we down-sample using a coarse-graining procedure. By applying these techniques, we obtain a new data set $D^k_{IJ}$, where $k$ runs now over a sparser set of frequencies $f^{k}_{IJ}$, with weights $w^k_{IJ}$ corresponding to the number of points we average over in the coarse-graining procedure. The down-sampled data set will have statistical properties similar to the ones of the $\bar{D}^\textmd{k}_{IJ}$ while being easier to handle numerically~\cite{Caprini:2019pxz, Flauger:2020qyi}. \\
    
It would seem natural to describe the data using a Gaussian likelihood of the form:
\begin{equation}
\ln \mathcal{L}_{\rm G} (\vec{\theta} | D^k_{IJ}) = -\frac{N_d}{2} \sum_{k} \sum_{I,J} w^k_{IJ} \left[ 1 - D^k_{IJ} / D^{\rm Th}_{IJ} (f_k, \vec{\theta}) \right]^2  \; ,
\end{equation}
where $\vec{\theta} = \{\vec{\theta}_s, \vec{\theta}_n\} $ is the vector of parameters (with $\vec{\theta}_s$, $\vec{\theta}_n$ being the signal and noise parameters, respectively), and $D^{\rm Th}_{IJ} (f_k, \vec{\theta}) = \Omega_{IJ, \ {\rm GW} } (f_k, \vec{\theta}) + \Omega_{IJ, \, n } (f_k, \vec{\theta})$ is the theoretical model for the data (with $\Omega_{IJ, \, {\rm GW} } (f_k, \vec{\theta})$ and $\Omega_{IJ, \ n} (f_k, \vec{\theta})$, being the signal and noise model, respectively). However, it is known~\cite{Bond:1998qg, Sievers:2002tq, WMAP:2003pyh, Hamimeche:2008ai} that this compressed likelihood does not account for the mild non-Gaussianity of the full likelihood, giving systematically biased results. This bias might be corrected by introducing a log-normal likelihood:
\begin{equation}
\ln \mathcal{L}_{\rm LN} (\vec{\theta} | D^k_{IJ}) = -\frac{N_d}{2} \sum_{k} \sum_{I,J} w^k_{IJ} \ln^2 \left[ D^{\rm Th}_{IJ} (f_k, \vec{\theta}) / D^k_{IJ}   \right] \; ,
\end{equation}
and considering the final likelihood to be:
\begin{equation}
\ln \mathcal{L} (\vec{\theta} ) = \frac{1}{3} \ln \mathcal{L}_{\rm G} (\vec{\theta} | D^k_{IJ}) +  \frac{2}{3} \ln \mathcal{L}_{\rm LN} (\vec{\theta} | D^k_{IJ}) \; .
\end{equation}
Consistently with the discussion in~\cref{sec:noise_single_link}, we include a prior for all the noise parameters in our analysis. These priors are chosen to be Gaussian, centered around the face values, i.e., $A_{ij}=3$, $P_{ij}=15$, $\forall \; \{ij\}$ combinations, and with a 20$\%$ width. Finally, to sample the parameter space we use the \texttt{emcee} sampler~\cite{Foreman-Mackey:2012any}, and results are visualized in contour plots using \texttt{Chainconsumer}~\cite{Hinton2016}.\\
    
For completeness, we report the expression of the SNR~\cite{Romano:2016dpx}:
\begin{equation}
\label{eq:SNR_def}
\textrm{SNR} \equiv \sqrt{T_d \; \sum_i \int \textrm{d}f \left( \frac{S_i^{\rm GW}}{S_i^{\rm N}} \right)^2 } \; , 
\end{equation}
where the index $i$ runs over all the TDI channels, which for simplicity, are assumed to be orthogonal.
    
\section{Noise parameter reconstruction for AET and AE\texorpdfstring{$\zeta$}{zeta}}%
\label{sec:noise_analysis}
In this section, we present the analysis of the noise parameter reconstruction, which is complementary to the discussion of~\cref{sec:parameter-reconstruction}. For this purpose, we start, in~\cref{sub:toymodel}, by considering a simple toy model, which provides useful insight for the interpretation of our results. We then proceed, in~\cref{sec:appendix_equal_noises} and~\cref{sec:appendix_unequal_noises}, with the discussion of the noise parameter reconstruction for the equal and unequal noise level cases, respectively.
    
\subsection{Toy model}%
\label{sub:toymodel}
Let us consider two time series, $D_1=N_1+N_2$ and $D_2=N_1$, where $N_1$ and $N_2$ are uncorrelated noises with PSDs $a$ and $b (f/f_\star)$ respectively, i.e., while the PSD of $N_1$ is a constant, that of $N_2$ has a frequency dependence.  The covariance $C_{ij} $, with $\{i,j\}\in \{1,2\}$, of the two data sets is given by:
\begin{equation}
C_{ij} \equiv  \mathrm{Cov}(D_i,D_j)=
\left(
\begin{array}{cc}
a+b\frac{f}{f_\star} & a \\
a & a
\end{array}
\right) \; .
\label{eq:toymodelcovDiDj}
\end{equation}
Assuming the data to be Gaussian, the log-likelihood reads:
\begin{equation}
\label{eq:Gaussian_likelihood}
- \log \mathcal{L} (D_i \vert a, b ) \propto \sum_f \left\{ \ln \left[ \det(C_{ij}) \right] + D_i  C^{-1}_{ij} D_j^* \right\}\; , 
\end{equation}
and it's easy to show that the Fisher matrix $F_{ij}$ can be computed as:
\begin{equation}
F_{ij} \equiv - \left. \frac{\partial^2 \log \mathcal{L} }{ \partial \theta_i \partial \theta_j } \right|_{\vec{\theta} =  \vec{\theta}_0} = \sum_f \Tr\left[ C^{-1} \frac{\partial C}{\partial \theta_i} C^{-1} \frac{\partial C}{\partial \theta_j} \right] \; , 
\end{equation}
where $\vec{\theta}_0$ are the best-fit parameters, which for~\cref{eq:Gaussian_likelihood} are given by $C_{ij } = D_i D^*_j$. To simplify the analysis, in the following, we replace the sum over finite frequencies with a continuous integral over the frequency range (going from some $f_{ \mathrm{min} }$ to $f_{ \mathrm{max} }$) multiplied by the total observation time $T$\footnote{Assuming the integration time is sufficiently long to have a reasonably fine sampling in $f$, the results in the two cases will coincide.}. From the equations above, it is easy to compute the covariance matrix (which is given by the inverse of the Fisher matrix) of the parameters $a$ and $b$. It reads: 
\begin{equation}
\label{eq:cov_ab}
\mathrm{Cov}(a,b)= \frac{1}{T ( f_{ \mathrm{max} } -f_{ \mathrm{min} } )}
\left(
\begin{array}{cc}
a^2 & 0 \\
0 & b^2
\end{array}
\right) \; .
\end{equation}
Since $D_2$ gives an independent readout of $N_1$, and $N_2$ can be estimated from $D_1 -D_2$, we can get independent estimates of the two parameters $a$ and $b$, whose variances scale with $a^2$ and $b^2$, respectively. To have better insight into the impact of the off-diagonal terms of~\cref{eq:toymodelcovDiDj} on~\cref{eq:cov_ab}, we repeat this calculation in the case in which these terms are neglected. Before entering into details, let us first note that if we consider a frequency range defined by $f_{\mathrm{min}}=0$ and $f_{\mathrm{max}}\rightarrow \infty$, the information diverges, the variance goes to zero, and we recover the above result. Let us proceed by considering $f_{\mathrm{max}}\gg f_{\mathrm{min}}$, with $b f_{\mathrm{max}}/f_\star \ll a$. In this case, the signal $D_2$ is subdominant over the entire frequency range and~\cref{eq:cov_ab} evaluates to:
\begin{equation}
\mathrm{Cov}(a,b)  \simeq \frac{1}{T f_{\mathrm{max}}}
\left(
\begin{array}{cc}
0.8\, a^2  & -1.2\,a^2 \frac{f_\star}{f_{\mathrm{max}}}  \\
-1.2\,a^2 \frac{f_\star}{f_{\mathrm{max}}}  & 4.8\,a^2 \frac{f_\star^2}{f_{\mathrm{max}}^2} 
\end{array}
\right)\; , 
\end{equation}
implying that the variance of $a$ is underestimated compared to the true value (which is equal to $a^2$), while that of $b$ depends on $a$ and is large compared to the true value (which is $b^2$). The second case we consider is $f_{\mathrm{max}}\gg f_{\mathrm{min}}$, with $b f_{\mathrm{max}}/f_\star = 2 a$. Then, the two signals are of similar amplitude, and~\cref{eq:cov_ab} reads:
\begin{equation}
\mathrm{Cov}(a,b) \simeq \frac{1}{T f_{\mathrm{max}}} 
\left(
\begin{array}{cc}
0.88\, a^2 &  -0.8\, ab \\
-0.8\, ab &  5\, b^2
\end{array}
\right) \;.
\end{equation}
In this case, the covariance of $a$ continues to be underestimated by a factor of order $4/5$, while that of $b$ is overestimated by a factor $\sim 5$. Finally, we consider $f_{\mathrm{max}}\gg f_{\mathrm{min}}$, with $b f_{\mathrm{max}}/f_\star \gg a$, so that $D_2$ is much larger than $D_1$ and~\cref{eq:cov_ab} reads: 
\begin{equation}
\mathrm{Cov}(a,b)  \simeq \frac{1}{T f_{\mathrm{max}}} 
\left(
\begin{array}{cc}
a^2   &  \frac{a^2 f_\star}{f_{\mathrm{max}}}\\
\frac{a^2f_\star}{f_{\mathrm{max}}} &  b^2
\end{array}
\right)\;.
\end{equation}
In this case, the frequency band is sufficiently broad such that either signal is clearly visible in at least part of it, and it is possible to recover the true covariance to a good approximation despite our having neglected the off-diagonal terms. Notice that in all three cases, neglecting off-diagonal terms in~\cref{eq:toymodelcovDiDj}, introduces degeneracies in the estimates of $a$ and $b$.\\
    
The findings of this toy model are consistent with the rationale that the off-diagonal elements add information that helps break degeneracies between model parameters, and that when they are neglected, the dominant signal, if present in multiple data streams, will have artificially more statistical weight and will thus be recovered with artificially small error bars, while the subdominant signal will be difficult to identify with much accuracy. 
    
\subsection{Equal noises} %
\label{sec:appendix_equal_noises}
 \begin{figure}[htbp!]
\centering
\includegraphics[width=\textwidth]{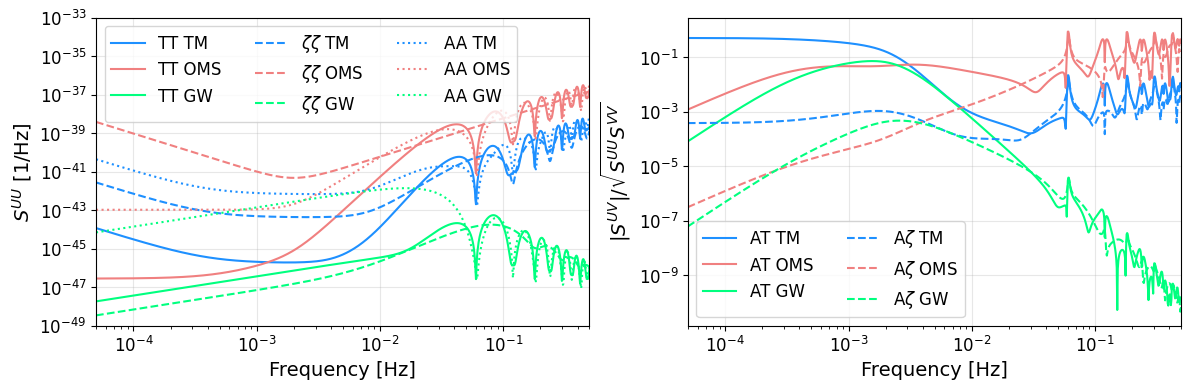}
\caption{\label{fig:equalnoisespectra} Left: TM, OMS, and GW PSDs in the case of unequal arms but equal noise spectra. Right: TM, OMS, and GW CSDs divided by the square root of the product of the PSDs for AA and TT or $\zeta\zeta$ (i.e., the square root of the component-wise coherence), in the case of unequal arms but equal noises.}
\end{figure}
    
In order to better understand the constraints on the noise parameters, we start by presenting in~\cref{fig:equalnoisespectra} the PSDs of the TM, OMS, and GW signal for the AA, TT, and $\zeta\zeta$ variables in the left-hand plot, and the CSDs of TM, OMS and GW divided by the square root of the product of the PSDs for AA and TT or $\zeta\zeta$ in the right-hand plot (i.e., the component-wise square-root of the coherence).  For example, ``AT TM'' is computed as $|S^{\mathrm{AT},\mathrm{TM}} | / \sqrt{ (S^{\mathrm{AA},\mathrm{N}}+ S^{\mathrm{AA},\mathrm{GW}})(S^{\mathrm{TT},\mathrm{N}}+ S^{\mathrm{TT},\mathrm{GW}})}$. Restricting the focus to the low-frequency range ($f<10^{-3}$ Hz), and to the variable T, we find that OMS is suppressed with respect to TM, reaching levels almost as small as GW. The TM cross-correlations in AT are at the 10\% level, while those for OMS and GW are several orders of magnitude smaller. Turning our attention to $\zeta$, we find that in the low-frequency range, contrary to T, the TM and OMS PSDs remains large compared to GW. Moreover, the cross-correlations for AE$\zeta$ are significantly smaller than those of AET. It is also worth noting that the hierarchy between the TM, OMS, and GW PSDs remains the same throughout the entire frequency range ($3\times 10^{-5}$ to 0.5 Hz) for $\zeta\zeta$, whereas the TM and OMS spectra intersect at $10^{-3}$ Hz for TT. The TM and OMS cross-correlations intersect at $f \simeq 5\times 10^{-3}$ Hz for both AT and A$\zeta$. Finally, at frequencies above $10^{-2}$ Hz, TM, OMS, and GW have comparable amplitudes in AET and AE$\zeta$. What differs is mainly the oscillatory features introduced by the TDI transfer functions. While the GW cross-correlations are strongly suppressed at large frequencies, the TM and OMS ones are much larger and of order 1\% and 10\% respectively. \\
    
\begin{figure}[htbp!]
\centering
\begin{subfigure}{0.485\textwidth}
\includegraphics[width=\textwidth]{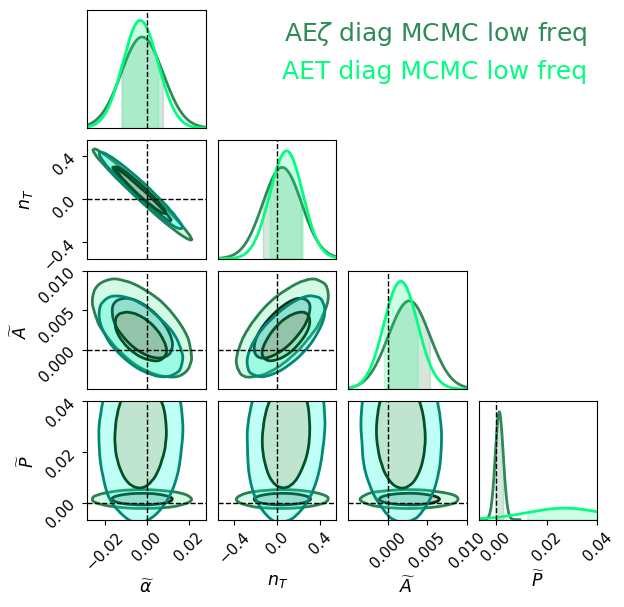}
\caption{Low frequency range.}
\label{fig:lowfreq2param}
\end{subfigure}
\hfill
\begin{subfigure}{0.485\textwidth}
\includegraphics[width=\textwidth]{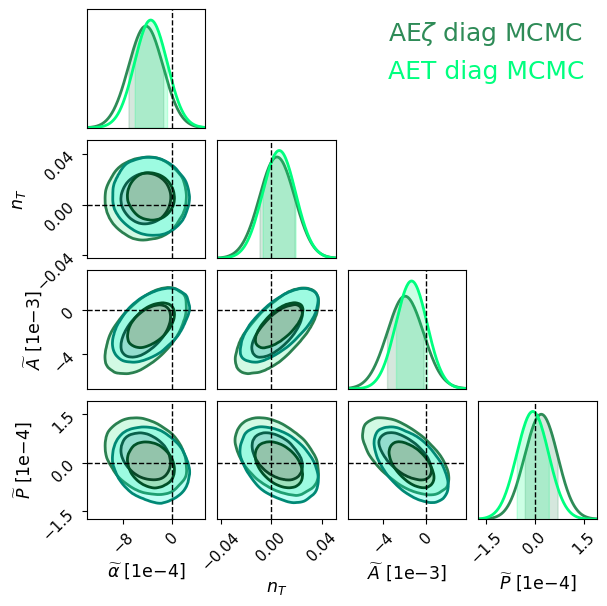}
\caption{Full frequency range.}
\label{fig:allfreq2param}
\end{subfigure}  
\caption{AET diag. MCMC vs. AE$\zeta$ diag. MCMC stochastic background and noise parameter reconstruction for unequal arms and equal noises.
\label{fig:contours2param}
}
\end{figure}

Let us proceed by discussing the parameter reconstruction in the case of equal noise levels, first restricting the analysis to the low-frequency ($f \leq 10^{-3}$Hz) range. The results of the MCMC runs for this configuration are presented in~\cref{fig:lowfreq2param}. As shown in~\cref{fig:equalnoisespectra}, at low frequencies, while the $AE\zeta$ basis is nearly orthogonal, correlations are relevant for AET.  Indeed, for AET, neglecting off-diagonal terms significantly impacts the posterior for the OMS amplitude (see the marginalized posterior for parameter $P$ in~\cref{fig:lowfreq2param}) which becomes largely underconstrained. This is due to the fact that OMS is strongly suppressed relative to TM in AET, such that much information could be gained from including the relatively large CSDs shown in the right-hand plot of~\cref{fig:equalnoisespectra}. The situation is in fact similar to that of $N_2$, and its associated parameter $b$, discussed in the toy model of~\cref{sub:toymodel}. On the other hand, the posteriors for all other parameters are slightly narrower than when considering the full AET matrix. This is once again consistent with what was found in~\cref{sub:toymodel}, for $N_1$ and associated parameter $a$. It is also worth noting that $\alpha$ and $n_T$ are strongly degenerate in both AET and AE$\zeta$. This happens in the low-frequency range, since the pivot frequency ($f_\star = \SI{3.873e-3}{\hertz}$) is located outside this range such that a change in $n_T$ induces a strong change in the GW signal's amplitude within the frequency range.  In this sense, including the pivot frequency in the frequency range considered, or widening the frequency range would break this degeneracy. This is indeed what happens once we consider the full frequency range, as we discuss below.\\
    
We conclude this section by including the higher frequency range (up to $5 \times 10^{-1}$Hz, consistent with the analysis in~\cref{sec:parameter-reconstruction}), which we expect will include significant information to the GW and noise parameters for both AET and AE$\zeta$. Indeed, the results shown in~\cref{fig:allfreq2param} show a great improvement in the determination of most parameters. In particular, the OMS noise parameter $P$ gains a factor of order 500 in the width of the marginalized posterior for AET, and a factor of order 50 for AE$\zeta$. This is consistent with the findings obtained in the example of~\cref{sub:toymodel}, where it was found that by extending the frequency range, one could capture additional information on sub-dominant components in the signal. Finally, comparing with~\cref{fig:allfreq2param}, we notice that including higher frequencies breaks the degeneracy between $\alpha$ and $n_T$ parameters. Indeed, as shown in~\cref{fig:equalnoisespectra}, the GW PSD grows at frequencies above $10^{-2}$ Hz, while the CSD drops significantly. While not explicitly shown in~\cref{fig:contours2param}, all these results are consistent with the corresponding Fisher matrix analyses.

\subsection{Unequal noises}%
\label{sec:appendix_unequal_noises}

\begin{figure}[htbp!]
\begin{center}
\includegraphics[width=\textwidth]{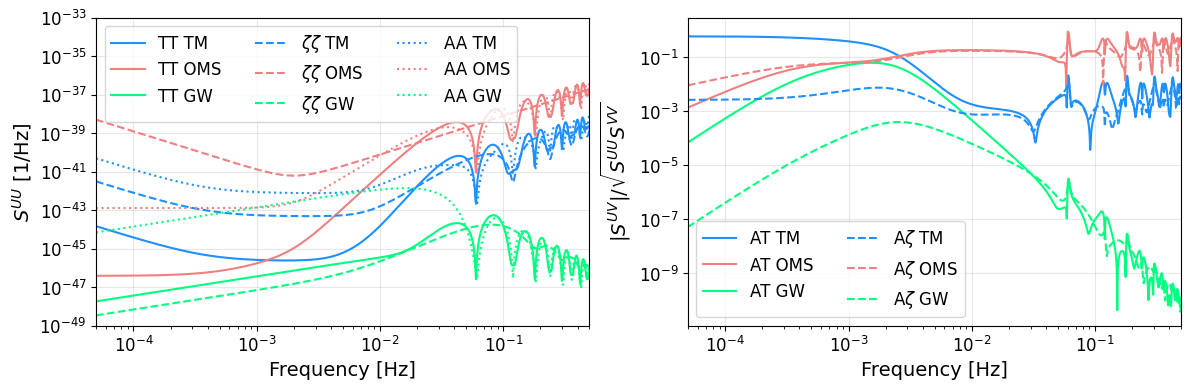}
\caption{\label{fig:unequalnoisespectra} Left: TM, OMS, and GW PSDs in the case of unequal arms and unequal noise spectra. Right: TM, OMS, and GW CSDs divided by the square root of the product of the PSDs for AA and TT or $\zeta\zeta$ (i.e., the square root of the component-wise coherence), in the case of unequal arms and unequal noise spectra.}
\end{center}
\end{figure}

In this section, we discuss the situation of most interest, the case of unequal arms and unequal noise levels. Comparing the left-hand plot of~\cref{fig:unequalnoisespectra} with the one of~\cref{fig:equalnoisespectra}, one finds that the TM, OMS, and GW PSDs in AA, TT, and $\zeta\zeta$, are practically unchanged with respect to the equal noise levels case. Turning attention to the right-hand plot of~\cref{fig:unequalnoisespectra}, one can make the following observations. Firstly, the OMS noise cross-correlation in A$\zeta$ is much larger than its counterpart in the equal noise case, over the entire frequency band. Indeed, this quantity is of the same order as the one for AT for frequencies $\gtrsim 5 \times 10^{-4}$Hz and is larger for lower frequencies. Secondly, the A$\zeta$ TM cross-correlations are also enhanced, by close to an order of magnitude in the band between $10^{-5}$ and $10^{-1}$ Hz. Thirdly, as expected, the GW cross-correlations are not affected by the introduction of independent noise levels and, contrarily to the equal noise level case, they stay smaller than the TM and OMS ones over the entire range. Finally, while the cross-correlations reach above the 10\% for both noises and for both TDI basis, the GW signal off-diagonal terms remain below 10\% for both noises and are significantly smaller for A$\zeta$. Drawing from these observations, one anticipates that the largest impact of neglecting off-diagonal terms will be seen in the OMS noise.\\

\begin{figure}[htbp!]
\centering
\begin{subfigure}{.495\textwidth}
\includegraphics[width=\textwidth]{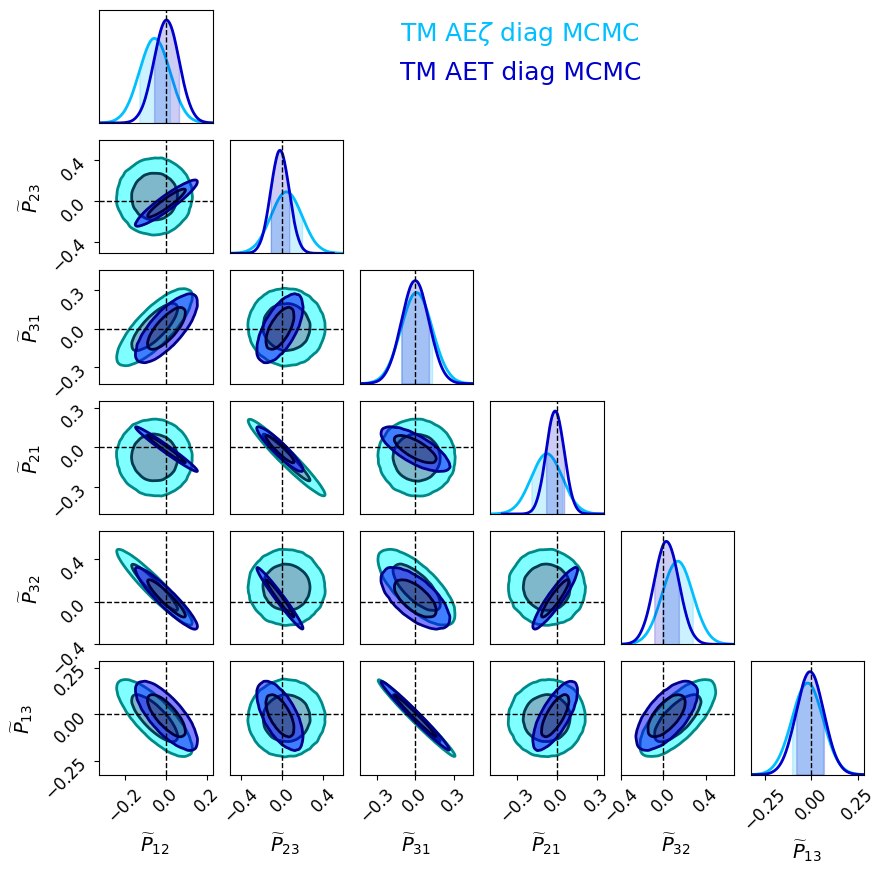}
\caption{Test mass (TM) acceleration noise}
\label{fig:fullfreq12param_TM}
\end{subfigure}
\begin{subfigure}{.495\textwidth}
\includegraphics[width=\textwidth]{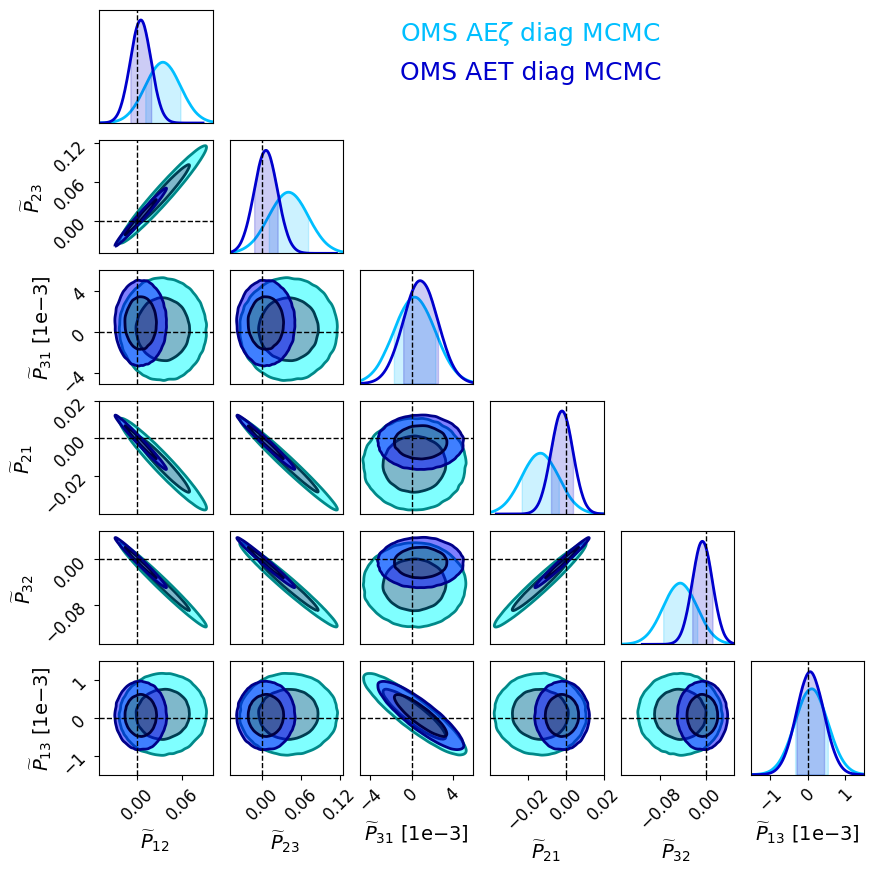}
\caption{Optical metrology system (OMS) noise }
\label{fig:fullfreq12param_OMS}
\end{subfigure}  
\caption{AET diag. MCMC vs. AE$\zeta$ diag. MCMC for the TM and OMS parameter posteriors obtained from the MCMC runs in the case of unequal arms, unequal noises, and over the full frequency range. \label{fig:fullfreq12param}}
\end{figure}

 \begin{figure}[htbp!]
\centering
\begin{subfigure}{\textwidth}
\includegraphics[width=0.9\textwidth]{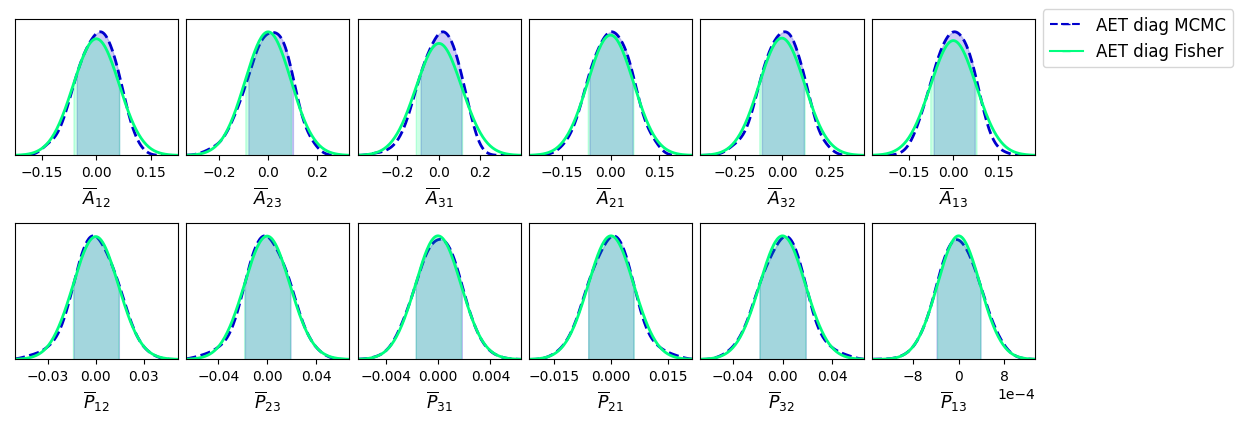}
\caption{AET diag. Fisher vs. MCMC diag.  TM on the first line and OMS on the second line.}
\label{fig:fullfreq12param_AET}
\end{subfigure}
\begin{subfigure}{\textwidth}
\includegraphics[width=0.9\textwidth]{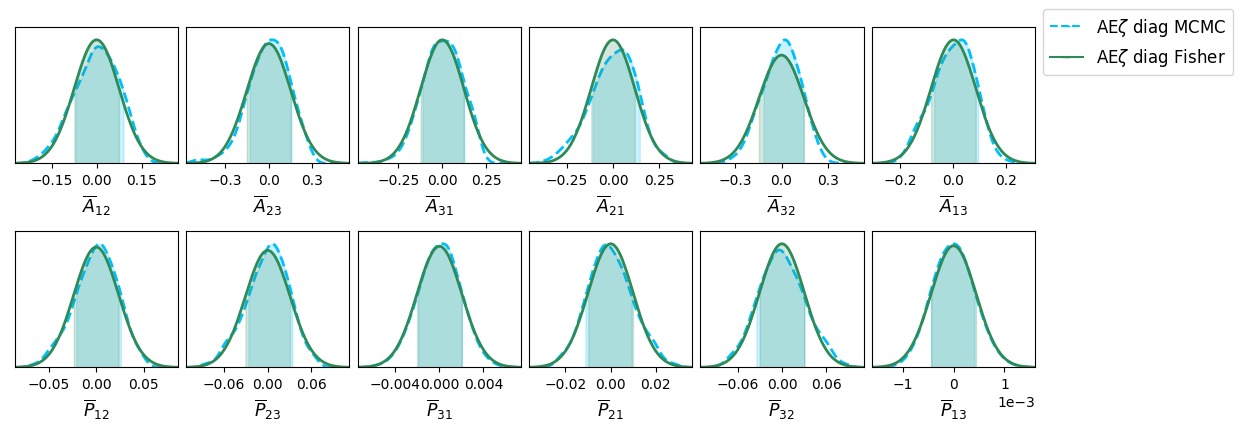}
\caption{AE$\zeta$ diag. Fisher vs. MCMC diag.  TM on the first line and OMS on the second line.}
\label{fig:fullfreq12param_AEZ}
\end{subfigure}
\caption{TM and OMS one-dimensional parameter posteriors obtained from Fisher forecasts and MCMC runs in the case of unequal arms and unequal noise levels.}
\end{figure}
 
Let us now focus on the AET and AE$\zeta$ MCMC results.  \Cref{fig:fullfreq12param_TM} and~\cref{fig:fullfreq12param_OMS} show the results of the AET and AE$\zeta$ runs for the TM and OMS parameters, respectively.  As shown in the right-hand plot of~\cref{fig:unequalnoisespectra}, the TM cross-correlations are significantly larger in AET than in AE$\zeta$ while the OMS cross-correlations are non-negligible and of similar amplitude in both AE$\zeta$ and AET.  Furthermore, as shown in the left-hand plot of that figure, the TM and OMS power spectral densities are significantly larger in $\zeta$ compared to T over a significant fraction of the frequency band. One can therefore conclude that similar to what was found in the toy model of~\cref{sub:toymodel}, ignoring cross-correlations will induce differences in the TM and OMS posterior widths for AET than in AE$\zeta$, and in particular, may lead to variance underestimation for AET.  This is true for 4 out of the 6 $A_{ij}$'s and 3 out of the 6 $P_{ij}$'s.  Let us recall that a Gaussian prior centered around the face values $A=3$, $P=15$, and with a standard deviation equal to the 20$\%$ of the central value was used for all the TM and OMS noise parameters. With this in mind, we note that, while all the $P_{ij}$ have posteriors that are considerably narrower than the priors, some of the $A_{ij}$ constraints are actually comparable in size, implying that the data fails to provide significant information on those parameters. Finally, we note that there exist degeneracies between the $A_{ij}$ and $A_{ji}$, and $P_{ij}$ and $P_{ji}$ pairs. This is expected because whether AET or AE$\zeta$, all measurements that involve the quantity $\eta_{ij}$ also involve its time-delayed counterpart $\eta_{ji}$. For what concerns the comparison between the AET and AE$\zeta$ results obtained via MCMC sampling and the FIM analysis assuming a diagonal TDI correlation matrix, while we do not present the full corner plots, we show in~\cref{fig:fullfreq12param_AET} and~\cref{fig:fullfreq12param_AEZ} the 1D posteriors for AET and AE$\zeta$, respectively. As is clear from these plots, the constraints obtained with the approaches agree. \\

\begin{figure}[htbp!]
\centering
\begin{subfigure}{.495\textwidth}
\includegraphics[width=\textwidth]{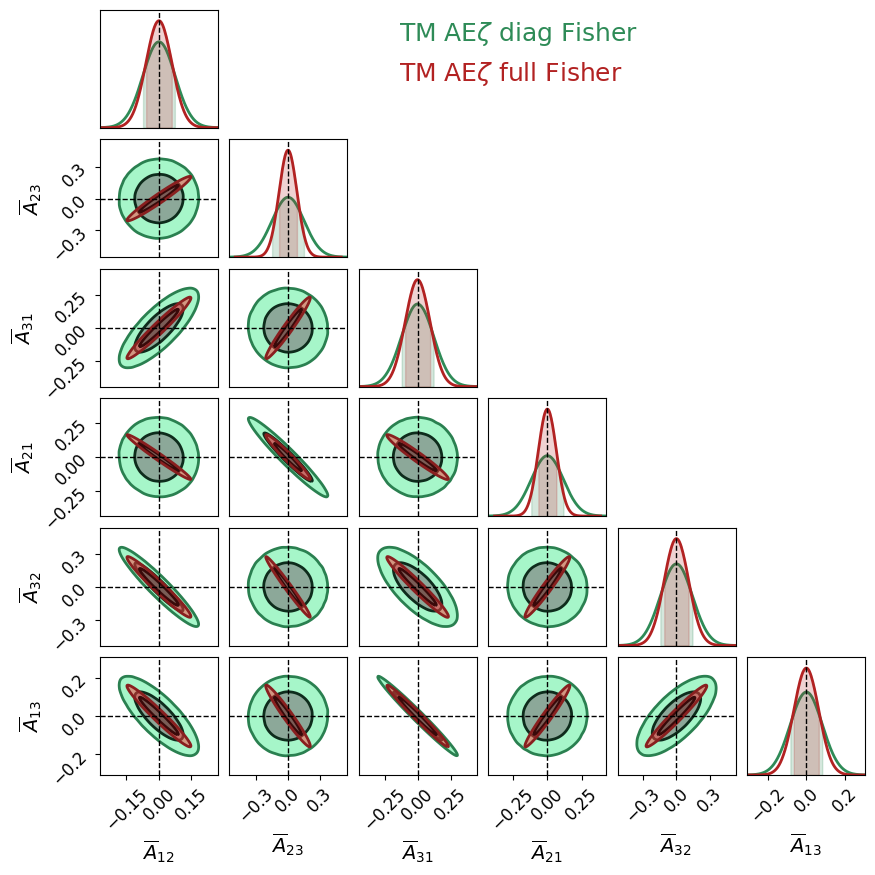}
\caption{TM acceleration noise.}
\label{fig:fullfreq12param_corr_TM}
\end{subfigure}
\begin{subfigure}{.495\textwidth}
\includegraphics[width=\textwidth]{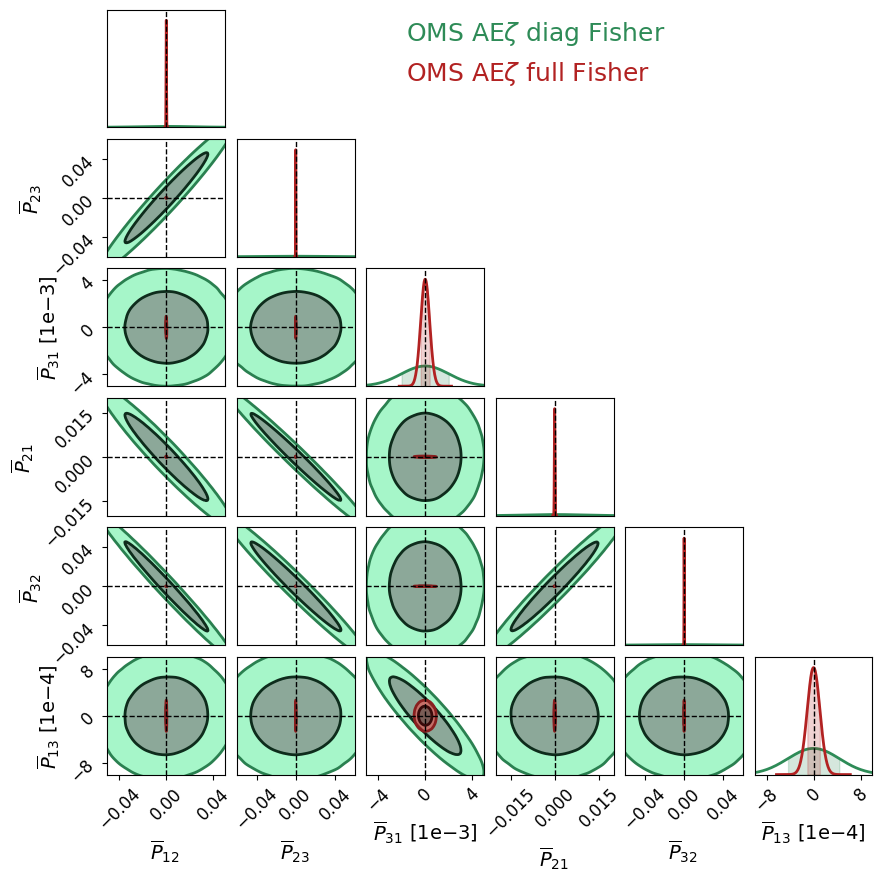}
\caption{OMS noise.}
\label{fig:fullfreq12param_corr_OMS}
\end{subfigure}
\caption{TM and OMS  posterior widths obtained from FIM analysis in the case of unequal arms and unequal noise levels when including or neglecting the off-diagonal terms in the TDI correlation matrix.}
\label{fig:fullfreq12param_corr}
\end{figure}

We conclude this appendix with a comparison of the Fisher analyses performed with the diagonal and full TDI covariance matrices. Since AET and AE$\zeta$ show slightly different, but qualitatively similar behaviours, we only show the corner plot for the latter, see~\cref{fig:unequalnoisespectra}.  Given that the AE$\zeta$ TM noise CSDs are relatively small, the constraints on the $A_{ij}$ parameters constraints are not expected to change significantly in going from a diagonal to a full correlation matrix analysis. This is verified in the 1D marginalized posteriors of~\cref{fig:fullfreq12param_corr_TM}. The introduction of correlations in the analysis does introduce correlations among parameters, with the result that the surface area of 2D contours shrinks when considering the full matrix. 
 Let us now turn our attention to OMS.  Given the CSDs of~\cref{fig:unequalnoisespectra} which for AE$\zeta$ (and also for AET), exhibit correlations that are greater than $10\%$ for a broad set of frequencies, we expect that including or excluding correlations of the TDI variable will significantly modify the constraints on the $P_{ij}$'s. Indeed, as illustrated in the toy model of~\cref{sub:toymodel}, ignoring cross-correlations means that the amplitude parameters of all subdominant noise components are largely unconstrained. This is confirmed by the contour plots of \cref{fig:fullfreq12param_corr_OMS}.

\FloatBarrier
\bibliography{references}
\end{document}